\documentclass[acmsmall,screen,nonacm]{acmart}

\usepackage{listings}
\usepackage{xcolor}
\usepackage{caption}
\usepackage[T1]{fontenc}
\usepackage{geometry}
\usepackage{tabularx,booktabs}
\usepackage{ragged2e} 
\usepackage{multirow}
\usepackage{subcaption}

\lstdefinelanguage{MLIR}
{
  keywords=[1]{
    addf,affine,apply,affine,
    alloc,br,cf,cmpi,cond_br,constant,dim,dma_start,dma_wait,
    for,func,load,memref,mulf,store,
    polyfor,tf,reduce,reshape,matmul,reduce_sum
  },
  keywordstyle=[1]\color{brown}\sffamily\bfseries,
  keywords=[2]{
    memref,tensor,vector,i1,i32,f32,f64,i64
  },
  keywordstyle=[2]\color{olive}\bfseries,
  keywords=[3]{
    copy,depths,schedule,unroll_jam,unroll_jam_factors,vectorize
  },
  keywordstyle=[3]\color{magenta},
  keywords=[4]{
    func, with, affine_map, affine_set, step
  },
  keywordstyle=[4]\color{orange}\bfseries,
  morecomment=[l]{//},
  commentstyle=\color{blue},
  columns=flexible,
  mathescape,
  tabsize=2,
  basicstyle=\scriptsize\ttfamily
}

\makeatletter
\newcommand{\halfpointsize}{\@setfontsize{\onepointsize}{0.5pt}{0.5pt}}
\makeatletter
\newcommand{\onepointsize}{\@setfontsize{\onepointsize}{1pt}{1pt}}
\makeatletter
\newcommand{\twopointsize}{\@setfontsize{\twopointsize}{2pt}{2pt}}
\makeatletter
\newcommand{\threepointsize}{\@setfontsize{\threepointsize}{3pt}{3pt}}
\makeatletter
\newcommand{\lsrcsize}{\@setfontsize{\lsrcsize}{4pt}{4pt}}
\makeatletter
\newcommand{\fourpointsize}{\@setfontsize{\fourpointsize}{4pt}{4pt}}
\makeatletter
\newcommand{\fivepointsize}{\@setfontsize{\fivepointsize}{5pt}{5pt}}

\newcommand{\Hilight}{\makebox[0pt][l]{\color{yellow}\rule[-4pt]{1.1\linewidth}{10pt}}}

\AtBeginDocument{%
  }

\setcopyright{acmlicensed}
\copyrightyear{2018}
\acmYear{2018}
\acmDOI{XXXXXXX.XXXXXXX}
\acmConference[Conference acronym 'XX]{Make sure to enter the correct
  conference title from your rights confirmation email}{June 03--05,
  2018}{Woodstock, NY}
\acmISBN{978-1-4503-XXXX-X/2018/06}

\begin{document}

\title{PolyBlocks: A Compiler Infrastructure for AI Chips and
Programming Frameworks}

\author{Uday Bondhugula}
\email{uday@polymagelabs.com}
\affiliation{%
  \institution{Polymage Labs and Indian Institute of Science}
  \city{Bengaluru}
  \state{Karnataka}
  \country{India}
}

\author{Akshay Baviskar}
\affiliation{%
  \institution{Polymage Labs}
  \city{Bengaluru}
  \country{India}}

\author{Navdeep Katel}
\affiliation{%
  \institution{Polymage Labs}
  \city{Bengaluru}
  \country{India}}

\author{Vimal Patel}
\affiliation{%
  \institution{Polymage Labs}
  \city{Bengaluru}
  \country{India}}

\author{Anoop JS}
\affiliation{%
  \institution{Polymage Labs}
  \city{Bengaluru}
  \country{India}}

\author{Arnab Dutta}
\affiliation{%
  \institution{Polymage Labs}
  \city{Bengaluru}
  \country{India}}

\renewcommand{\shortauthors}{Bondhugula et al.}

\begin{abstract}
  We present the design and implementation of PolyBlocks, a modular and reusable
  MLIR-based compiler infrastructure for AI programming frameworks and AI chips.
  PolyBlocks is based on pass pipelines that compose transformations on loop nests
  and SSA, primarily relying on lightweight affine access analysis; the
  transformations are stitched together in specialized ways to realize
  high-performance code automatically by the use of analytical cost models and
  heuristics. The optimizations in these passes include multi-level tiling,
  fusion, on-chip scratchpad usage, mapping matmuls and convolutions to matrix
  units, fusing the attention layer, and several other transformations for
  parallelism and locality. They have been developed in a way that makes it easy
  to build PolyBlocks-based compilers to target new chips, reusing much of the
  infrastructure. PolyBlocks' design and architecture enables fully automatic code
  generation from high-level frameworks to low-level target-specific intrinsics.

  Experimental results from evaluating PolyBlocks-powered just-in-time compilation for
  PyTorch and JAX targeting NVIDIA GPUs show that it is able to match or
  outperform Torch Inductor and XLA in several cases, although the latter rely
  on a combination of vendor libraries and code generation. For individual
  operators like matmuls and convolutions, PolyBlocks-generated code is competitive
  with the best vendor-tuned libraries or hand-written kernels.

\end{abstract}





\maketitle

\section{Introduction}
\label{sec:intro}
Compilers are software systems that translate high-level program statements into
instructions that the hardware can execute. With programming frameworks developing
 with higher-level abstractions and hardware chips with specialized
features, compilers bridge larger gaps. For the domain of artificial
intelligence (AI), there has been a rapid innovation in building specialized
accelerator chips, with the desire to have high-level programming frameworks such
as PyTorch~\cite{pytorch,pytorch-github} and JAX~\cite{jax} supported on them, while
exploiting as much performance from the hardware as possible.

Tensorflow~\cite{tensorflow2015-whitepaper},
PyTorch~\cite{pytorch,pytorch-github}, and JAX~\cite{jax} are among the three
dominant high-level frameworks to express ML and AI models.  The standard way to
execute these models is called {\it eager} execution. This involves executing the
graph of tensor operators one after another, which the {\it executor} of
the respective framework does by calling a pre-compiled and pre-optimized
library routine available for the operator. A majority of high performance
realized today from the execution of AI models comes from highly-optimized
vendor libraries that the eager executors rely on.  These include libraries
such as MKL, OpenBLAS, CuDNN~\cite{chetlur14arxiv}, CuBLAS, ROCm,
FlashAttention~\cite{dao22arxiv}, and TensorRT-LLM~\cite{tensorrt}.  When
developers are not satisfied with the performance delivered by these libraries
on GPUs, re-implementing the desired operators or a group of them with low-level
abstractions like CUDA, Cutlass~\cite{thakkar2023cutlass},
Triton~\cite{triton19tillet}, C/C++ or inline assembly is common, requiring
deeper optimization expertise and knowledge of the hardware.

{\bf Eager vs compiled:} Besides the {\it eager} mode of execution
described above, high-level frameworks like PyTorch, JAX, and Tensorflow also
provide the {\it compiled} mode through other backends, often called the
JIT (just-in-time) compiled mode. In the compiled mode, the graph of operators
is captured, optimized, and processed as a whole, and code is generated for its
various parts to a large extent. {\it Fusion} of operators is one of the
greatest strengths of the compiled approach, reducing the number of trips made
to the global memory, and in some cases even eliminating accesses to on-chip
memory altogether. This greatly speeds up memory-bandwidth-bound computation,
which is beneficial for nearly any chip. In addition, it also speeds up
compositions of compute-bound operations (like matmuls, convolutions) with
dependent memory bandwidth operations. Modern AI accelerators often include
specialized hardware in the form of matrix units or tensor cores to dramatically
speed up matrix-matrix multiplications. With Amdahl's law, the balance tilts
making optimization of other memory-bandwidth-bound computation important.

The PyTorch, JAX, and Tensorflow frameworks are served by different compilers
that are effective separately for each. While TensorFlow and JAX have had
XLA~\cite{xla}, Torch Inductor~\cite{pytorch24asplos} has been reported to be
the most effective for PyTorch~\cite{pytorch24asplos} in terms of coverage and
performance. These compilers do not share infrastructure beyond the lowest
layers when targeting CPUs or GPUs. Supporting targets very different to CPUs
or GPUs involves significant additional effort due to these compilers' use of
multiple different intermediate representations (IR) and their reliance on
vendor libraries to realize performance for key operators, especially on CPUs
and GPUs.

{\bf Programming landscape: high vs mid vs low-level frameworks:} Although not
standard terminology, we frequently use the {\it high, mid, and low} taxonomy in
this paper to better describe the programming landscape for AI chips today.
Popular programming approaches to develop and execute deep learning models fall
into the following three categories: (1) low-level ones that include CUDA,
OpenCL, C/C++ with abstractions such as Cutlass~\cite{thakkar2023cutlass}, (2)
mid-level ones based on tile-level abstractions such as
Triton~\cite{triton19tillet,triton-web} and Pallas/Mosaic~\cite{pallas}, and (3)
high-level ones like PyTorch, JAX, and Tensorflow.
Figure~\ref{fig:high_mid_low} illustrates how productivity and abstraction
change across these. Compilers are employed for all three classes of frameworks,
but the challenges of building them are vastly different.  Compilation for
PyTorch or JAX is more challenging than for a framework like Triton, with the
former requiring more work in lowering abstractions.

\definecolor{codegreen}{rgb}{0,0.6,0}
\definecolor{codegray}{rgb}{0.9,0.9,0.9}
\definecolor{codepurple}{rgb}{0.58,0,0.82}
\definecolor{tritongreen}{rgb}{0.9,1,0.9} 
\definecolor{backcolour}{rgb}{0.95,0.95,0.95} 

\lstdefinestyle{tritonstyle}{
    language=Python,
    backgroundcolor=\color{codegray},
    keywordstyle=\color{magenta},
    stringstyle=\color{codepurple},
    breakatwhitespace=false,
    breaklines=true,
    keepspaces=true,
    showspaces=false,
    showstringspaces=false,
    showtabs=false,
    tabsize=2,
    frame=single,
    framerule=0pt,
}

\noindent 
\begin{figure}[htbp]
\begin{minipage}[t]{0.31\textwidth}
\begin{lstlisting}[language=python,basicstyle=\ttfamily\tiny,title={PyTorch}]
# Torch function to matrix matrix
# multiply A and B. Compilation
# can be turned on or off.
@torch.compile
def matmul(A, B):
    return torch.matmul(A, B)
\end{lstlisting}
\end{minipage}%
\hfill%
\begin{minipage}[t]{0.44\textwidth}
\begin{lstlisting}[style=tritonstyle,basicstyle=\ttfamily\fourpointsize,title={Triton}]
# triton.jit'ed functions can be auto-tuned using the `triton.autotune`
# decorator.
@triton.autotune(
    configs=get_autotune_config(),
    key=['M', 'N', 'K'],
)
@triton.jit
def matmul_kernel(
    # Pointers to matrices
    a_ptr, b_ptr, c_ptr,
    # Matrix dimensions
    M, N, K,
    # The stride variables represent how much to increase the ptr by when moving by 1
    # element in a particular dimension. E.g. 'stride_am' is how much to increase 'a_ptr'
    # by to get the element one row down (A has M rows).
    stride_am, stride_ak,
    stride_bk, stride_bn,
    stride_cm, stride_cn,
    # Meta-parameters
    BLOCK_SIZE_M: tl.constexpr, BLOCK_SIZE_N: tl.constexpr, BLOCK_SIZE_K: tl.constexpr,
    GROUP_SIZE_M: tl.constexpr,
):
    """Kernel for computing the matmul C = A x B.
    A has shape (M, K), B has shape (K, N) and C has shape (M, N)
    """
    # Map program ids 'pid' to the block of C it should compute.
    # This is done in a grouped ordering to promote L2 data reuse.
    # See above 'L2 Cache Optimizations' section for details.
    pid = tl.program_id(axis=0)
    num_pid_m = tl.cdiv(M, BLOCK_SIZE_M)
    num_pid_n = tl.cdiv(N, BLOCK_SIZE_N)
    num_pid_in_group = GROUP_SIZE_M * num_pid_n
    group_id = pid // num_pid_in_group
    first_pid_m = group_id * GROUP_SIZE_M
    group_size = min(num_pid_m - first_pid_m, GROUP_SIZE_M)
    pid_m = first_pid_m + (pid % group_size)
    pid_n = (pid % num_pid_in_group) // group_size

    # Create pointers for the first blocks of A and B.
    # We will advance this pointer as we move in the K direction
    # and accumulate
    # 'a_ptrs' is a block of [BLOCK_SIZE_M, BLOCK_SIZE_K] pointers
    # 'b_ptrs' is a block of [BLOCK_SIZE_K, BLOCK_SIZE_N] pointers
    # See above 'Pointer Arithmetics' section for details
    offs_am = (pid_m * BLOCK_SIZE_M + tl.arange(0, BLOCK_SIZE_M)) % M
    offs_bn = (pid_n * BLOCK_SIZE_N + tl.arange(0, BLOCK_SIZE_N)) % N
    offs_k = tl.arange(0, BLOCK_SIZE_K)
    a_ptrs = a_ptr + (offs_am[:, None] * stride_am + offs_k[None, :] * stride_ak)
    b_ptrs = b_ptr + (offs_k[:, None] * stride_bk + offs_bn[None, :] * stride_bn)

    # Iterate to compute a block of the C matrix.
    # We accumulate into a [BLOCK_SIZE_M, BLOCK_SIZE_N] block
    # of fp32 values for higher precision.
    # 'accumulator' will be converted back to fp16 after the loop
    accumulator = tl.zeros((BLOCK_SIZE_M, BLOCK_SIZE_N), dtype=tl.float32)
    for k in range(0, tl.cdiv(K, BLOCK_SIZE_K)):
        # Load the next block of A and B, generate a mask by checking the K dimension.
        # If it is out of bounds, set it to 0.
        a = tl.load(a_ptrs, mask=offs_k[None, :] < K - k * BLOCK_SIZE_K, other=0.0)
        b = tl.load(b_ptrs, mask=offs_k[:, None] < K - k * BLOCK_SIZE_K, other=0.0)
        # We accumulate along the K dimension.
        accumulator += tl.dot(a, b)
        # Advance the pointers
        a_ptrs += BLOCK_SIZE_K * stride_ak
        b_ptrs += BLOCK_SIZE_K * stride_bk

    # Write back the block of the output matrix C with masks.
    offs_cm = pid_m * BLOCK_SIZE_M + tl.arange(0, BLOCK_SIZE_M)
    offs_cn = pid_n * BLOCK_SIZE_N + tl.arange(0, BLOCK_SIZE_N)
    c_ptrs = c_ptr + stride_cm * offs_cm[:, None] + stride_cn * offs_cn[None, :]
    mask = (offs_cm[:, None] < M) & (offs_cn[None, :] < N)
    tl.store(c_ptrs, c, mask=mask)
\end{lstlisting}
\end{minipage}%
\hfill%
\begin{minipage}[t]{0.11\textwidth}
\begin{lstlisting}[language=c++,basicstyle=\ttfamily\onepointsize,title=CUDA]
template <
    typename T, size_t BLOCK_TILE_SIZE_X, size_t BLOCK_TILE_SIZE_Y,
    size_t BLOCK_TILE_SIZE_K, size_t WARP_TILE_SIZE_X, size_t WARP_TILE_SIZE_Y,
    size_t WMMA_TILE_SIZE_X, size_t WMMA_TILE_SIZE_Y, size_t WMMA_TILE_SIZE_K,
    size_t NUM_WMMA_TILES_X, size_t NUM_WMMA_TILES_Y, size_t NUM_WMMA_TILES_K,
    size_t BLOCK_TILE_SKEW_SIZE_X, size_t BLOCK_TILE_SKEW_SIZE_Y>
__device__ void process_data_from_shared_memory_using_wmma(
    nvcuda::wmma::fragment<nvcuda::wmma::matrix_a, WMMA_TILE_SIZE_Y,
                           WMMA_TILE_SIZE_X, WMMA_TILE_SIZE_K, T,
                           nvcuda::wmma::col_major>
        a_frags[NUM_WMMA_TILES_Y],
    nvcuda::wmma::fragment<nvcuda::wmma::matrix_b, WMMA_TILE_SIZE_Y,
                           WMMA_TILE_SIZE_X, WMMA_TILE_SIZE_K, T,
                           nvcuda::wmma::row_major>
        b_frags[NUM_WMMA_TILES_X],
    nvcuda::wmma::fragment<nvcuda::wmma::accumulator, WMMA_TILE_SIZE_Y,
                           WMMA_TILE_SIZE_X, WMMA_TILE_SIZE_K, T>
        acc_frags[NUM_WMMA_TILES_Y][NUM_WMMA_TILES_X],
    T const A_thread_block_tile_transposed[BLOCK_TILE_SIZE_K]
                                          [BLOCK_TILE_SIZE_Y +
                                           BLOCK_TILE_SKEW_SIZE_Y],
    T const B_thread_block_tile[BLOCK_TILE_SIZE_K]
                               [BLOCK_TILE_SIZE_X + BLOCK_TILE_SKEW_SIZE_X],
    size_t warp_row_idx, size_t warp_col_idx) {
#pragma unroll
  for (size_t k_i{0U}; k_i < NUM_WMMA_TILES_K; ++k_i) {
#pragma unroll
    for (size_t wmma_tile_row_idx{0U}; wmma_tile_row_idx < NUM_WMMA_TILES_Y;
         ++wmma_tile_row_idx) {
      nvcuda::wmma::load_matrix_sync(
          a_frags[wmma_tile_row_idx],
          &A_thread_block_tile_transposed[k_i * WMMA_TILE_SIZE_K]
                                         [warp_row_idx * WARP_TILE_SIZE_Y +
                                          wmma_tile_row_idx * WMMA_TILE_SIZE_Y],
          BLOCK_TILE_SIZE_Y + BLOCK_TILE_SKEW_SIZE_Y);
    }
#pragma unroll
    for (size_t wmma_tile_col_idx{0U}; wmma_tile_col_idx < NUM_WMMA_TILES_X;
         ++wmma_tile_col_idx) {
      nvcuda::wmma::load_matrix_sync(
          b_frags[wmma_tile_col_idx],
          &B_thread_block_tile[k_i * WMMA_TILE_SIZE_K]
                              [warp_col_idx * WARP_TILE_SIZE_X +
                               wmma_tile_col_idx * WMMA_TILE_SIZE_X],
          BLOCK_TILE_SIZE_X + BLOCK_TILE_SKEW_SIZE_X);
    }
#pragma unroll
    for (size_t wmma_tile_row_idx{0U}; wmma_tile_row_idx < NUM_WMMA_TILES_Y;
         ++wmma_tile_row_idx) {
#pragma unroll
      for (size_t wmma_tile_col_idx{0U}; wmma_tile_col_idx < NUM_WMMA_TILES_X;
           ++wmma_tile_col_idx) {
        // Perform the matrix multiplication.
        nvcuda::wmma::mma_sync(acc_frags[wmma_tile_row_idx][wmma_tile_col_idx],
                               a_frags[wmma_tile_row_idx],
                               b_frags[wmma_tile_col_idx],
                               acc_frags[wmma_tile_row_idx][wmma_tile_col_idx]);
      }
    }
  }
}

// GEMM kernel v07.
// Each thread in the block processes THREAD_TILE_SIZE_Y *
// THREAD_TILE_SIZE_X output values. Number of threads BLOCK_TILE_SIZE_Y *
// BLOCK_TILE_SIZE_X / (THREAD_TILE_SIZE_Y * THREAD_TILE_SIZE_X)
template <typename T, size_t BLOCK_TILE_SIZE_X, size_t BLOCK_TILE_SIZE_Y,
          size_t BLOCK_TILE_SIZE_K, size_t BLOCK_TILE_SKEW_SIZE_X,
          size_t BLOCK_TILE_SKEW_SIZE_Y, size_t WARP_TILE_SIZE_X,
          size_t WARP_TILE_SIZE_Y, size_t WMMA_TILE_SIZE_X,
          size_t WMMA_TILE_SIZE_Y, size_t WMMA_TILE_SIZE_K, size_t NUM_THREADS>
__global__ void
gemm_v07_vectorized_double_buffered(size_t m, size_t n, size_t k, T alpha,
                                    T const *A, size_t lda, T const *B,
                                    size_t ldb, T beta, T *C, size_t ldc) {
  constexpr size_t NUM_WARPS_X{BLOCK_TILE_SIZE_X / WARP_TILE_SIZE_X};
  constexpr size_t NUM_WARPS_Y{BLOCK_TILE_SIZE_Y / WARP_TILE_SIZE_Y};
  static_assert(BLOCK_TILE_SIZE_X % WARP_TILE_SIZE_X == 0U);
  static_assert(BLOCK_TILE_SIZE_Y % WARP_TILE_SIZE_Y == 0U);

  constexpr size_t NUM_WMMA_TILES_X{WARP_TILE_SIZE_X / WMMA_TILE_SIZE_X};
  static_assert(WARP_TILE_SIZE_X % WMMA_TILE_SIZE_X == 0U);
  constexpr size_t NUM_WMMA_TILES_Y{WARP_TILE_SIZE_Y / WMMA_TILE_SIZE_Y};
  static_assert(WARP_TILE_SIZE_Y % WMMA_TILE_SIZE_Y == 0U);
  constexpr size_t NUM_WMMA_TILES_K{BLOCK_TILE_SIZE_K / WMMA_TILE_SIZE_K};
  static_assert(BLOCK_TILE_SIZE_K % WMMA_TILE_SIZE_K == 0U);

  constexpr size_t NUM_PIPELINES{2U};
  // Only double buffer is supported in the implementation.
  // But even more number of pipelines can be supported if the implementation
  // is modified.
  static_assert(NUM_PIPELINES == 2U);
  static_assert((NUM_WARPS_X * NUM_WARPS_Y) % NUM_PIPELINES == 0U);
  static_assert(NUM_THREADS % NUM_PIPELINES == 0U);
  constexpr size_t NUM_THREADS_PER_PIPELINE{NUM_THREADS / NUM_PIPELINES};
  constexpr size_t NUM_WARPS_PER_PIPELINE{(NUM_WARPS_X * NUM_WARPS_Y) /
                                          NUM_PIPELINES};

  // Cache a tile of A and B in shared memory for data reuse.
  __shared__ T A_thread_block_tile_transposed[NUM_PIPELINES][BLOCK_TILE_SIZE_K]
                                             [BLOCK_TILE_SIZE_Y +
                                              BLOCK_TILE_SKEW_SIZE_Y];
  __shared__ T B_thread_block_tile[NUM_PIPELINES][BLOCK_TILE_SIZE_K]
                                  [BLOCK_TILE_SIZE_X + BLOCK_TILE_SKEW_SIZE_X];

  // Declare the fragments.
  nvcuda::wmma::fragment<nvcuda::wmma::matrix_a, WMMA_TILE_SIZE_Y,
                         WMMA_TILE_SIZE_X, WMMA_TILE_SIZE_K, T,
                         nvcuda::wmma::col_major>
      a_frags[NUM_WMMA_TILES_Y];
  nvcuda::wmma::fragment<nvcuda::wmma::matrix_b, WMMA_TILE_SIZE_Y,
                         WMMA_TILE_SIZE_X, WMMA_TILE_SIZE_K, T,
                         nvcuda::wmma::row_major>
      b_frags[NUM_WMMA_TILES_X];
  nvcuda::wmma::fragment<nvcuda::wmma::accumulator, WMMA_TILE_SIZE_Y,
                         WMMA_TILE_SIZE_X, WMMA_TILE_SIZE_K, T>
      acc_frags[NUM_WMMA_TILES_Y][NUM_WMMA_TILES_X];
  nvcuda::wmma::fragment<nvcuda::wmma::accumulator, WMMA_TILE_SIZE_Y,
                         WMMA_TILE_SIZE_X, WMMA_TILE_SIZE_K, T>
      c_frag;

// Make sure the accumulator starts from 0.
#pragma unroll
  for (size_t wmma_tile_row_idx{0U}; wmma_tile_row_idx < NUM_WMMA_TILES_Y;
       ++wmma_tile_row_idx) {
    for (size_t wmma_tile_col_idx{0U}; wmma_tile_col_idx < NUM_WMMA_TILES_X;
         ++wmma_tile_col_idx) {
      nvcuda::wmma::fill_fragment(
          acc_frags[wmma_tile_row_idx][wmma_tile_col_idx], static_cast<T>(0));
    }
  }

  size_t const thread_linear_idx{threadIdx.y * blockDim.x + threadIdx.x};
  size_t const warp_linear_idx{thread_linear_idx / 32U};
  size_t const warp_row_idx{warp_linear_idx / NUM_WARPS_X};
  size_t const warp_col_idx{warp_linear_idx % NUM_WARPS_X};
  // Separate the warps to different pipelines.
  size_t const pipeline_index{warp_linear_idx / NUM_WARPS_PER_PIPELINE};

	// Number of outer loops to perform the sum of inner products.
  // C_thread_block_tile =
  // \sigma_{thread_block_tile_idx=0}^{num_thread_block_tiles-1} A[:,
  // thread_block_tile_idx:BLOCK_TILE_SIZE_K] *
  // B[thread_block_tile_idx:BLOCK_TILE_SIZE_K, :]
  size_t const num_thread_block_tiles{(k + BLOCK_TILE_SIZE_K - 1) /
                                      BLOCK_TILE_SIZE_K};

  if (pipeline_index == 0U) {
    // Pipeline 0 warps load buffer 0.
    load_data_from_global_memory_to_shared_memory_transposed_vectorized<
        T, BLOCK_TILE_SIZE_X, BLOCK_TILE_SIZE_Y, BLOCK_TILE_SIZE_K,
        NUM_THREADS_PER_PIPELINE, BLOCK_TILE_SKEW_SIZE_X,
        BLOCK_TILE_SKEW_SIZE_Y>(
        A, lda, B, ldb, A_thread_block_tile_transposed[pipeline_index],
        B_thread_block_tile[pipeline_index], 0U,
        thread_linear_idx - pipeline_index * NUM_THREADS_PER_PIPELINE, m, n, k);
  }
  __syncthreads();

  // Continued...
\end{lstlisting}
\end{minipage}
\begin{minipage}[t]{0.11\textwidth}
\begin{lstlisting}[language=c++,basicstyle=\ttfamily\onepointsize,title={\hspace{0.2in}}]
  for (size_t thread_block_tile_idx{0U};
       thread_block_tile_idx < num_thread_block_tiles;
       thread_block_tile_idx += NUM_PIPELINES) {
    if (pipeline_index == 0U) {
      // Pipeline 0 warps process buffer 0.
      process_data_from_shared_memory_using_wmma<
          T, BLOCK_TILE_SIZE_X, BLOCK_TILE_SIZE_Y, BLOCK_TILE_SIZE_K,
          WARP_TILE_SIZE_X, WARP_TILE_SIZE_Y, WMMA_TILE_SIZE_X,
          WMMA_TILE_SIZE_Y, WMMA_TILE_SIZE_K, NUM_WMMA_TILES_X,
          NUM_WMMA_TILES_Y, NUM_WMMA_TILES_K, BLOCK_TILE_SKEW_SIZE_X,
          BLOCK_TILE_SKEW_SIZE_Y>(
          a_frags, b_frags, acc_frags,
          A_thread_block_tile_transposed[pipeline_index],
          B_thread_block_tile[pipeline_index], warp_row_idx, warp_col_idx);
      __syncthreads();

      // Pipeline 0 warps process buffer 1.
      if (thread_block_tile_idx + 1U < num_thread_block_tiles) {
        process_data_from_shared_memory_using_wmma<
            T, BLOCK_TILE_SIZE_X, BLOCK_TILE_SIZE_Y, BLOCK_TILE_SIZE_K,
            WARP_TILE_SIZE_X, WARP_TILE_SIZE_Y, WMMA_TILE_SIZE_X,
            WMMA_TILE_SIZE_Y, WMMA_TILE_SIZE_K, NUM_WMMA_TILES_X,
            NUM_WMMA_TILES_Y, NUM_WMMA_TILES_K, BLOCK_TILE_SKEW_SIZE_X,
            BLOCK_TILE_SKEW_SIZE_Y>(
            a_frags, b_frags, acc_frags,
            A_thread_block_tile_transposed[pipeline_index + 1],
            B_thread_block_tile[pipeline_index + 1], warp_row_idx,
            warp_col_idx);
      }
      __syncthreads();

      // Pipeline 0 warps load buffer 0.
      if (thread_block_tile_idx + 2U < num_thread_block_tiles) {
        load_data_from_global_memory_to_shared_memory_transposed_vectorized<
            T, BLOCK_TILE_SIZE_X, BLOCK_TILE_SIZE_Y, BLOCK_TILE_SIZE_K,
            NUM_THREADS_PER_PIPELINE, BLOCK_TILE_SKEW_SIZE_X,
            BLOCK_TILE_SKEW_SIZE_Y>(
            A, lda, B, ldb, A_thread_block_tile_transposed[pipeline_index],
            B_thread_block_tile[pipeline_index], thread_block_tile_idx + 2,
            thread_linear_idx - pipeline_index * NUM_THREADS_PER_PIPELINE, m, n,
            k);
      }
      __syncthreads();
    } else {
      // Pipeline 1 warps load buffer 1.
      if (thread_block_tile_idx + 1U < num_thread_block_tiles) {
        load_data_from_global_memory_to_shared_memory_transposed_vectorized<
            T, BLOCK_TILE_SIZE_X, BLOCK_TILE_SIZE_Y, BLOCK_TILE_SIZE_K,
            NUM_THREADS_PER_PIPELINE, BLOCK_TILE_SKEW_SIZE_X,
            BLOCK_TILE_SKEW_SIZE_Y>(
            A, lda, B, ldb, A_thread_block_tile_transposed[pipeline_index],
            B_thread_block_tile[pipeline_index], thread_block_tile_idx + 1,
            thread_linear_idx - pipeline_index * NUM_THREADS_PER_PIPELINE, m, n,
            k);
      }
      __syncthreads();

      // Pipeline 1 warps process buffer 0.
      process_data_from_shared_memory_using_wmma<
          T, BLOCK_TILE_SIZE_X, BLOCK_TILE_SIZE_Y, BLOCK_TILE_SIZE_K,
          WARP_TILE_SIZE_X, WARP_TILE_SIZE_Y, WMMA_TILE_SIZE_X,
          WMMA_TILE_SIZE_Y, WMMA_TILE_SIZE_K, NUM_WMMA_TILES_X,
          NUM_WMMA_TILES_Y, NUM_WMMA_TILES_K, BLOCK_TILE_SKEW_SIZE_X,
          BLOCK_TILE_SKEW_SIZE_Y>(
          a_frags, b_frags, acc_frags,
          A_thread_block_tile_transposed[pipeline_index - 1],
          B_thread_block_tile[pipeline_index - 1], warp_row_idx, warp_col_idx);
      __syncthreads();

      // Pipeline 1 warps process buffer 1.
      if (thread_block_tile_idx + 1U < num_thread_block_tiles) {
        process_data_from_shared_memory_using_wmma<
            T, BLOCK_TILE_SIZE_X, BLOCK_TILE_SIZE_Y, BLOCK_TILE_SIZE_K,
            WARP_TILE_SIZE_X, WARP_TILE_SIZE_Y, WMMA_TILE_SIZE_X,
            WMMA_TILE_SIZE_Y, WMMA_TILE_SIZE_K, NUM_WMMA_TILES_X,
            NUM_WMMA_TILES_Y, NUM_WMMA_TILES_K, BLOCK_TILE_SKEW_SIZE_X,
            BLOCK_TILE_SKEW_SIZE_Y>(
            a_frags, b_frags, acc_frags,
            A_thread_block_tile_transposed[pipeline_index],
            B_thread_block_tile[pipeline_index], warp_row_idx, warp_col_idx);
      }
      __syncthreads();
    }
  }

// Write the results to DRAM.
#pragma unroll
  for (size_t wmma_tile_row_idx{0U}; wmma_tile_row_idx < NUM_WMMA_TILES_Y;
       ++wmma_tile_row_idx) {
#pragma unroll
    for (size_t wmma_tile_col_idx{0U}; wmma_tile_col_idx < NUM_WMMA_TILES_X;
         ++wmma_tile_col_idx) {
      // Load the fragment from global memory.
      nvcuda::wmma::load_matrix_sync(
          c_frag,
          &C[(blockIdx.y * BLOCK_TILE_SIZE_Y + warp_row_idx * WARP_TILE_SIZE_Y +
              wmma_tile_row_idx * WMMA_TILE_SIZE_Y) *
                 n +
             blockIdx.x * BLOCK_TILE_SIZE_X + warp_col_idx * WARP_TILE_SIZE_X +
             wmma_tile_col_idx * WMMA_TILE_SIZE_X],
          n, nvcuda::wmma::mem_row_major);
      // Perform scaling and addition.
      for (size_t i{0}; i < c_frag.num_elements; ++i) {
        c_frag.x[i] =
            alpha * acc_frags[wmma_tile_row_idx][wmma_tile_col_idx].x[i] +
            beta * c_frag.x[i];
      }
      // Store the fragment back to global memory.
      nvcuda::wmma::store_matrix_sync(
          &C[(blockIdx.y * BLOCK_TILE_SIZE_Y + warp_row_idx * WARP_TILE_SIZE_Y +
              wmma_tile_row_idx * WMMA_TILE_SIZE_Y) *
                 n +
             blockIdx.x * BLOCK_TILE_SIZE_X + warp_col_idx * WARP_TILE_SIZE_X +
             wmma_tile_col_idx * WMMA_TILE_SIZE_X],
          c_frag, n, nvcuda::wmma::mem_row_major);
    }
  }
}

template <typename T>
void launch_gemm_kernel_v07_vectorized_double_buffered(
    size_t m, size_t n, size_t k, T const* alpha, T const* A, size_t lda,
    T const* B, size_t ldb, T const* beta, T* C, size_t ldc,
    cudaStream_t stream)
{
    // Feel free to play with the block tile sizes.
    // The algorithm correctness should always be guaranteed.
    constexpr unsigned int BLOCK_TILE_SIZE_X{128U};
    constexpr unsigned int BLOCK_TILE_SIZE_Y{128U};
    constexpr unsigned int BLOCK_TILE_SIZE_K{16U};

    // The skew size is used to avoid bank conflicts in shared memory.
    constexpr size_t BLOCK_TILE_SKEW_SIZE_X{16U};
    constexpr size_t BLOCK_TILE_SKEW_SIZE_Y{16U};

    constexpr unsigned int WARP_TILE_SIZE_X{32U};
    constexpr unsigned int WARP_TILE_SIZE_Y{64U};
    constexpr unsigned int NUM_WARPS_X{BLOCK_TILE_SIZE_X / WARP_TILE_SIZE_X};
    constexpr unsigned int NUM_WARPS_Y{BLOCK_TILE_SIZE_Y / WARP_TILE_SIZE_Y};
    static_assert(BLOCK_TILE_SIZE_X % WARP_TILE_SIZE_X == 0U);
    static_assert(BLOCK_TILE_SIZE_Y % WARP_TILE_SIZE_Y == 0U);

    constexpr unsigned int WMMA_TILE_SIZE_X{16U};
    constexpr unsigned int WMMA_TILE_SIZE_Y{16U};
    constexpr unsigned int WMMA_TILE_SIZE_K{16U};

    constexpr unsigned int NUM_THREADS_PER_BLOCK{NUM_WARPS_X * NUM_WARPS_Y *
                                                 32U};

    dim3 const block_dim{NUM_THREADS_PER_BLOCK, 1U, 1U};
    dim3 const grid_dim{
        (static_cast<unsigned int>(n) + BLOCK_TILE_SIZE_X - 1U) /
            BLOCK_TILE_SIZE_X,
        (static_cast<unsigned int>(m) + BLOCK_TILE_SIZE_Y - 1U) /
            BLOCK_TILE_SIZE_Y,
        1U};
    gemm_v07_vectorized_double_buffered<
        T, BLOCK_TILE_SIZE_X, BLOCK_TILE_SIZE_Y, BLOCK_TILE_SIZE_K,
        BLOCK_TILE_SKEW_SIZE_X, BLOCK_TILE_SKEW_SIZE_Y, WARP_TILE_SIZE_X,
        WARP_TILE_SIZE_Y, WMMA_TILE_SIZE_X, WMMA_TILE_SIZE_Y, WMMA_TILE_SIZE_K,
        NUM_THREADS_PER_BLOCK><<<grid_dim, block_dim, 0U, stream>>>(
        m, n, k, *alpha, A, lda, B, ldb, *beta, C, ldc);
    CHECK_LAST_CUDA_ERROR();
}
\end{lstlisting}
\end{minipage}
\vskip -6pt
\caption{An indicative figure showing the trade-offs between high, mid, and low-level
programming frameworks while delivering comparable performance. Matmul with
PyTorch, Triton, and CUDA, all of which deliver close to peak
performance. CUDA courtesy: Lei Mao~\cite{cudaoptgemmleimao}.\label{fig:high_mid_low}}
\end{figure}



\textbf{Reliance on manually developed libraries vs fully code-generating:} While
Inductor, XLA, and TensorRT~\cite{tensorrt} are considered compilers, they rely on
libraries for the most compute-intensive operators, sometimes heavily, when
targeting CPUs and GPUs. For GPUs, they rely on CuDNN~\cite{chetlur14arxiv},
CuBLAS, FlashAttention~\cite{dao22arxiv}, and several other manually developed
kernels. The opposite of this approach is a {\it fully code-generating} one,
where the compiler generates all low-level code that is ultimately executed,
where this low-level code is typically in the form of instructions and
intrinsics for the hardware available in LLVM~\cite{llvm-project} or another
low-level IR. The choice to use existing libraries may often be dictated by
pragmatics since there has been no or little evidence of a high-level compiler
generating code competitive with CuDNN (for convolutions), or CuBLAS (for
matmuls), or kernels for the attention~\cite{vaswani17nips} layer.  The
automatic transformation and optimization ability needed to realize that level
of performance in a reusable way involves substantial effort. In addition,
significant effort may have already been put into building such libraries, and
layering on top of such libraries provides a separation of concerns. However,
there are also major downsides to using libraries in the long term: (1) they
become non-portable due to various versions and strategies needed for different
problem sizes and targets, choices that could have been encoded into a fully
code generating compiler; (2) the lack of good performance for certain cases
would leave a compiler built on top of such libraries with no choice, (3) fusion
of operators will not be possible with such libraries unless the libraries
provide all the desired compositions, something not practical, and (4) effort
going into libraries is often not reusable across different targets.

In this paper, we propose PolyBlocks, a new compiler infrastructure based on the
MLIR IR infrastructure~\cite{mlir21cgo,mlir-web}. The building blocks in PolyBlocks
are designed to support a {\it fully code-generating approach}, i.e., not
relying on any pre-developed libraries. We show that PolyBlocks' mid-level
transformations of fusion and tiling are more powerful than those of existing
compilers, leading to better cross-operator optimization. We show how
convolutions and matmuls are mapped to matrix units through compositions of
simpler transformations in a generic way on affine loop nests, while not
precluding tiled fusion. Similarly, we also achieve automatic fusion of
operators in the attention layer~\cite{vaswani17nips} computation through a
sequence of transformations. Even for individual operators' code generation,
code generated by PolyBlocks is often competitive with well-established low-level
and mid-level optimization approaches and even vendor-tuned libraries on NVIDIA
GPUs. More importantly, the PolyBlocks infrastructure allows one to quickly build
compilers for new AI chips without the need to redevelop several advanced
transformations.  When used as a backend for PyTorch, JAX, or TensorFlow for
GPUs, it provides significant improvements in several cases over
state-of-the-art approaches that rely on both libraries and code generation.

The remainder of this paper is organized as follows. In
Section~\ref{sec:design-choices}, we describe the various design choices we
made for PolyBlocks. Section~\ref{sec:architecture} describes the high-level
architecture of PolyBlocks. Section~\ref{sec:opts} describes in detail some of the key
optimizations available in PolyBlocks. Experimental evaluation and
performance comparison are provided in Section~\ref{sec:evaluation}. Related
work and conclusions are presented in Section~\ref{sec:related-work} and
Section~\ref{sec:conclusions} respectively.

\section{Design Choices}
\label{sec:design-choices}
Building a compiler infrastructure that spans the distance between high-level
frameworks and specialized chips presents multiple choices in various
steps. In this section, we discuss the design choices we made and their
rationale. Our objective here is to build an infrastructure for high-level
frameworks that provides an abstraction similar to that of TensorFlow, PyTorch, and JAX.

The presence of multiple levels of parallelism, matrix units or tensor cores,
vector units, and on-chip memory or scratchpads of various kinds that require
explicit data movement are common features of various specialized accelerators.
The ability to overlap computation and data movement is also important. The gap
between global memory bandwidth and compute speed also necessitates
cross-operator fusion to maximize reuse in on-chip memory and registers.


\textbf{Full automation first:}
There have been various approaches to compiling deep learning computations, with
different IRs, with programmer-supplied schedules, and reliance on auto-tuning.
With schedule-driven approaches, the optimization of models was guided by a
schedule of transformations separate from the specification. The schedule would
be different for not only different hardware but also different inputs. This
process requires experts who are knowledgeable about both compute patterns
and the target hardware throughout the optimization process.  In some way,
humans  become part of the compiler. We deliberately avoided this approach, as
it is not sustainable or scalable when dealing with the cross-product of
evolving models and hardware.  Our focus is on full automation to start with.
However, this choice does not preclude us from tuning or controlling various
parameters in the cost model for future purposes if input is desired. We also
wished to avoid relying on auto-tuning early on during the development to
restrict tuning to a very focused space and to reach that space through the
maturity of the compiler infrastructure and the cost models used.

\textbf{Fully code-generating approach:} For the reasons related to fusion,
maintainability, scalability, and extensibility to new hardware that were
discussed in Section~\ref{sec:intro}, we consciously avoided relying on
hand-written libraries, closed or open source, by choosing to build a fully
code-generating compiler all the way to target-specific intrinsics available.
This allows the developed passes and transformations to be more reusable each
time new hardware is to be supported, since newer hardware may not have
libraries as well-developed or mature as for the NVIDIA GPU ecosystem today.


\textbf{Reusable pass pipeline-based approach:} All of the above leads us to an
approach where we built a number of transformation passes for various key
pieces.  Before the availability of the MLIR~\cite{mlir-web,mlir21cgo}
intermediate representation (IR) infrastructure in 2019, we believe that the
basic toolkit to build compilers to traverse the entire abstraction distance
from high-level representations to generate code for hardware target intrinsics
was not available. In order to maximize reusable infrastructure, we structured
it as a sequence of MLIR pass pipelines, through which all IR, regardless of the input
framework, go. Nearly the same pipelines are used across different
hardware up to the mid-level stages.  More detail on this is provided in the
next section.

\textbf{Incremental programmability: one-line compile directive:}
Our desire was to make compilation equivalent to turning a switch on or off. Usability
would thus be similar to that of models like OpenMP. Both Torch
Inductor~\cite{pytorch24asplos} and XLA JIT~\cite{xla} provide such a user
experience, which is highly productive, with the eager mode co-existing for
initial debugging, reference, and validation purposes. As shown in
Figure~\ref{fig:polyblocks-model}, compilation is enabled by a one-line directive
on the desired function, either through an annotation or an API call. With this
approach, unmodified PyTorch, JAX, or Tensorflow models can be compiled and
executed.

\begin{figure}[!htbp]
  \begin{subfigure}[b]{0.48\linewidth}
    \centering
		\begin{lstlisting}[language=python,basicstyle=\ttfamily\tiny,escapechar=\%]
%\Hilight%@polyblocks_jit_torch(compile_options={'target': 'nvgpu'})
def blur_x_blur_y(img):
    # We are doing a regular image convolution in the
    # X direction and then in the Y direction.
    blur_kernel = kernel.view(1, 1, 1, 5)
    img_reshaped = img.view(3, 1, height, width)
    # The output is two short at each end.
    blurx = torch.nn.functional.conv2d(img_reshaped,
                         blur_kernel, stride=1, padding=0)
    blur_kernel = kernel.view(1, 1, 5, 1)
    blury = torch.nn.functional.conv2d(blurx, blur_kernel,
                                      stride=1, padding=0)
    output = blury.view(3, height - 4, width - 4)
    return output


out = blur_x_blur_y(img)
    \end{lstlisting}
    \caption{
      3x of Torch Inductor on NVIDIA A10.}
  \end{subfigure}%
  \hfill
  \begin{subfigure}[b]{0.48\linewidth}
  \begin{lstlisting}[language=python,basicstyle=\ttfamily\tiny,escapechar=\%]
    def attention(self, q, k, v, bias):
        k_t = torch.transpose(k, -1, -2)
        qk = torch.matmul(q, k_t)
        qk = qk + bias
        soft_qk = torch.nn.functional.softmax(qk,
                      dim = -1)
        res = torch.matmul(soft_qk, v)
        return res




    compiled = torch.compile(attention,
                  backend="polyblocks",
                  options={"target": "nvgpu"}
    )
      \end{lstlisting}
    \caption{PolyBlocks is registered as a Torch backend.}
  \end{subfigure}
  \vskip -5pt
  \caption{PyTorch functions JIT compiled with PolyBlocks. Incremental programmability.
  \label{fig:polyblocks-model}
  }
  \vskip -10pt
\end{figure}

\textbf{Reliance on built-in MLIR dialects and type system:}
We found the MLIR type system sufficient for all the core transformations,
including target-specific mid-level transformations. We consciously avoided
adding new dialects or operations for such purposes. These choices allowed
optimizations and lowerings to be available to a large class of higher-level
operations that were lowered to affine nests, and a resilience to input lowered
from different programming frameworks. We make use of external attributes over
short distances in the pass pipeline to convey information and hints to
heuristics and to avoid recomputation of certain properties
(Appendix~\ref{sec:polyblocks-kinds}).

\textbf{Use of affine or polyhedral techniques only where needed:} Polyhedral
techniques~\cite{isl,cloog,uday08cc,ppcg} are powerful in resolving a large
number of analyses arising from mid-level transformations such as fusion, tiling,
and generation of data movement code between fast and slow memories.  However,
they also rely on relatively expensive integer set operations,
which are often an overkill for many analyses when dealing with IR lowered from
tensor operators. A large number of cases are resolved with simple linear or
constant time checks on affine functions, and one has to defer to more expensive
polyhedral operations in a small set of cases. We also note that several of the
models that we compile and report performance for, or desire to scale to have up
to tens of thousands of loop nests (hundreds or a few thousands of named tensor
operators at the level of torch aten, mhlo, or stablehlo dialects). To make
compiler development practical, we believe that compile times have to be kept to the order
of a few seconds to a few tens of seconds, even for the largest models we see today.

\section{Architecture}
\label{sec:architecture}
In this section, we describe the architecture of the PolyBlocks core
engine and the template architecture of any compiler built with PolyBlocks.  The
reader is referred to the literature on MLIR~\cite{mlir-web}, \cite{mlir21cgo}
since the rest of this section, in various places, assumes familiarity with the
MLIR concepts of dialects, operations, built-in types, and its use of SSA.

\subsection{Choice of dialects and lowering paths}
\begin{figure}[htbp]
\centering
  \includegraphics[width=0.6\linewidth]{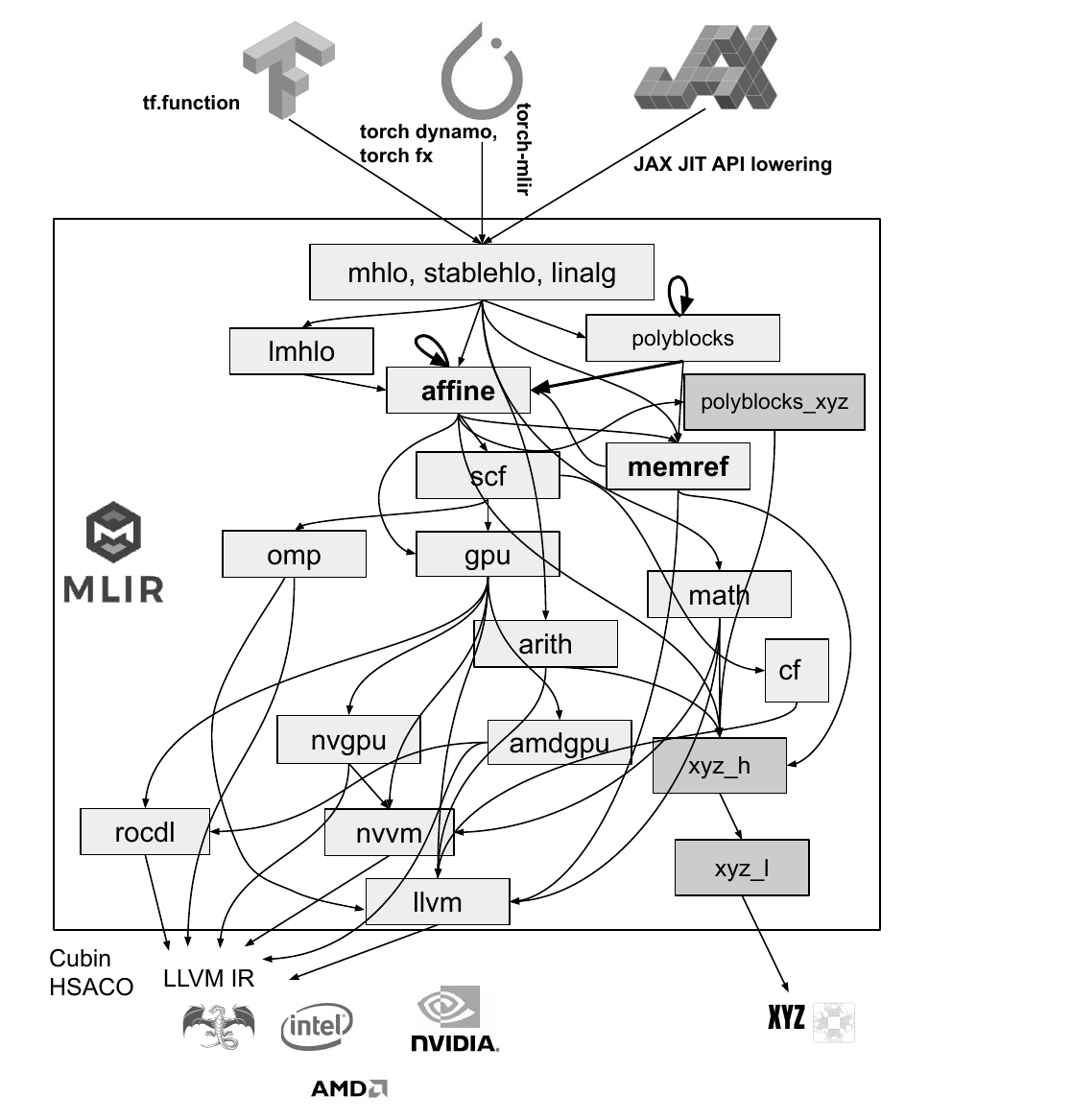}
  \vskip -5pt
	\caption{PolyBlocks compiler engine powering multiple programming frameworks and
hardware. MLIR dialect lowering paths and how new targets are supported (XYZ in
this figure).\label{fig:polyblocks-stack}}
\end{figure}

Figure~\ref{fig:polyblocks-stack} shows the lowering paths through dialects. It is
not meant to be fully exhaustive, but close to capturing lowering paths through
edges. All three frameworks, TensorFlow, PyTorch, and JAX, have had entry paths into
MLIR~\cite{mlir21cgo,mlir-web} for several years maintained by open-source
community-driven projects. The Tensorflow project includes a customizable pass
pipeline that lowers from the {\it tf} dialect to the {\it mhlo} dialect after
tf.functions are imported into the {\it tf} dialect. For JAX, the JAX JIT API
provides a way to lower JAX graphs to the {\it stablehlo} dialect. For PyTorch,
Torch dynamo is used to capture FX graphs, which can then be lowered by the
{\it torch-mlir}~\cite{torch-mlir} project: from the {\it torch} dialect through
{\it aten} to {\it linalg}. The {\it mhlo}, {\it stablehlo}, and {\it
linalg} dialects thus form the entry point into the PolyBlocks compiler engine.
Any combination of these dialects can be handled seamlessly.  All of these dialects
operator on MLIR {\it tensor} types and thus tensor value semantics are used.
Tensors are SSA values and can only be defined once. On these lowering paths
until the mix of {\it mhlo}, {\it linalg}, and {\it stablehlo}, the PolyBlocks
compiler driver uses additional customization whenever an alternate lowering is
desirable. In such cases, the {\it polyblocks} dialect is used to add a path for
such operations.

\subsection{A five-stage pass pipeline}
\label{sec:polyblocks-stages}
\begin{figure}[htbp]
  \centering
  \includegraphics[width=0.75\linewidth]{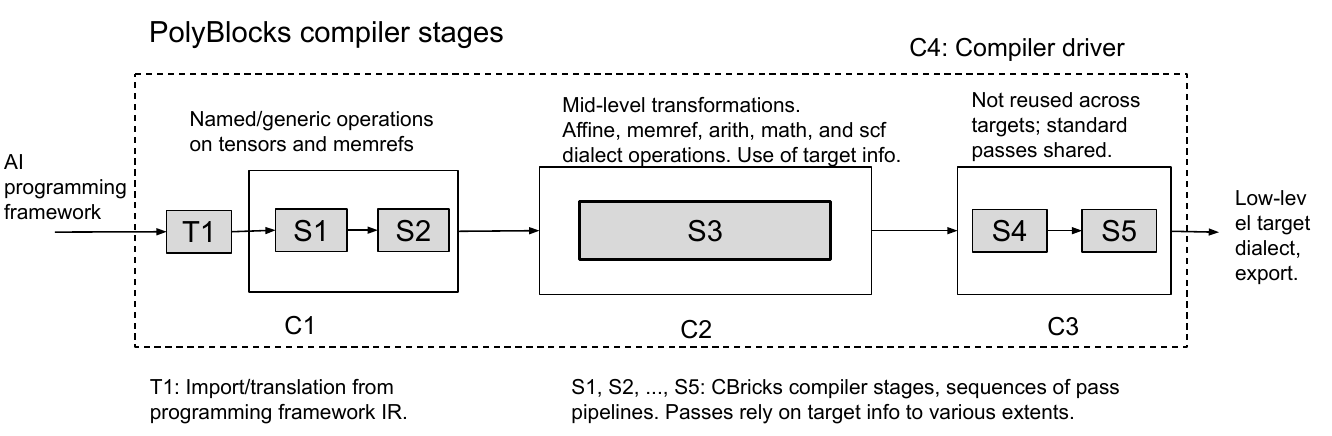}
  \vskip -10pt
  \caption{Overall components and high-level architecture of a compiler built
  from PolyBlocks.\label{fig:polyblocks-components}}
\vskip -5pt
\end{figure}

The PolyBlocks lowering and transformation infrastructure is organized as a five-stage
pass pipeline shown in Figure~\ref{fig:polyblocks-components}: S1 through S5 for
proper isolation, organization, and easier debugging purposes.

\textbf{Frontend stage:} The S1 and S2 stages are nearly target neutral. The S1
stage lowers from operations on {\it tensor} types to those on {\it memref}
types, i.e., buffer semantics are introduced. The S2 stage lowers from named ops
on {\it memref} types to affine loop nests. Very little target-specific
information is used by S1 and S2, and is often restricted to supplying
information on precision modes (e.g., using mixed-precision) or providing
supported matmul precision modes to specific passes, which may need this
information where lowering or expanding named ops.

\textbf{Mid-level optimizer:} The S3 stage is the mid-level optimizer where the
majority optimizations and mappings are performed, relying on target
information to various extents. The  S3 pass pipeline comprises 50-70 passes
typically for a target. It operates primarily on the {\it
affine}~\cite{mlir-affine} and {\it memref} dialect with the ability to realize
specialized code for a target, while still using MLIR's in-built type system.
Fusion, tiling, mapping to matrix units, generation of data movement code for
on-chip memories, and vectorization are performed in S3.  Standard dialects from
MLIR like {\it arith}, {\it math}, {\it builtin}, and {\it func} are used
throughout in most stages. Operations from the {\it vector} dialect are used as
a result of the vectorization.

\textbf{Backend stages:}
The S4 stage converts from the {\it affine} and {\it memref} form to the
{\it scf} dialect in conjunction with other target-specific dialects where
special abstractions are needed to represent parallelism and synchronization. For
GPUs, this is the {\it gpu} dialect, and the {\it omp} dialect for CPU
parallelization. The S5 stage then lowers the {\it scf} and other
target-specific parallelism abstractions out of MLIR. For NVIDIA GPUs, this would
lower to LLVM + NVVM through the {\it nvgpu} dialect (or
through the {\it amdgpu} dialect for AMD GPUs), and to LLVM for CPUs.
Component C3, which includes S4 and S5 is very target-specific and these are not
reused across targets beyond standard upstream passes like canonicalization,
dead code elimination, CSE, and invariant code motion.

\textbf{Compiler driver:}
Finally, all three components, C1, C2, and C3, are combined by a compiler
driver (C4), which is developed in Python. C4 defines the pass pipeline
specifications for all the different stages, passing IR through these different
stages, providing target information, and providing other debugging
capabilities. C4 includes the support to convert the graph representations of
the programming frameworks' to MLIR, leveraging existing open-source
infrastructure (T1). C4 crucially also contains the logic to process input and
output data when performing JIT compilation, i.e., converting between the programming
frameworks' tensor representation to the memref ABI that the compiled MLIR uses.
Support for ahead-of-time compilation is also handled here, where a binary or a
library can be generated for the specification being supplied. MLIR's LLVM
ORC-based~\cite{orcv2} execution engine is used to execute the compiled code to
realize the JIT for both CPUs and GPUs while linking the desired runtime
libraries. For other targets, available custom runtime paths and their Python
bindings are used to create the execution driver in C4. For NVIDIA GPUs, MLIR's
LLVM with NVVM intrinsics are translated through LLVM's optimization pipeline
through to its NVPTX backend and subsequently to GPU assembly through the CUDA
driver API. The CUDA runtime is the only dependency for the executable code for
NVGPUs and the ROCm runtime similarly for AMD GPUs.  For CPUs, the OpenMP
runtime is dynamically linked, while LLVM's OpenMP support is leveraged for the
generated code.

\textbf{Building a new PolyBlocks-based compiler:}
Due to the above organization, we have the following powerful properties and
well-isolated separation when creating new target backends for PolyBlocks.
Building a new compiler with PolyBlocks requires: (1) almost reusing everything as
is from T1, S1, and S2, (2) customizing the passes of S3 for the new target, (3)
building out S4 and S5 from scratch (reusing standard canonicalization,
low-level optimization passes), and (4) support in C4 for input/output,
kernel launching, and synchronization.

\section{Optimizations}
\label{sec:opts}
In this section, we describe some of the key components of the
mid-level optimizer, and how we built or derived passes based on upstream MLIR
infrastructure. As mentioned in the previous section, the dialects in bold in
Figure~\ref{fig:polyblocks-stack} is where most of the transformation and
optimization described in the next section is performed.  The use of {\it
affine} and {\it memref} dialects allows us to perform transformations in a very
generic and powerful way, resilient to changing patterns in the input
programming frameworks. Different kinds of convolutions, matmuls, and tensor
contractions, which might have been lowered from different kinds of ops and
differently named ops in programming frameworks, are all optimized to the same
extent, as our results and performance improvements will later demonstrate.  All
code snippets shown in this section are from PolyBlocks-generated IR.

\subsection{Use of affine analysis for various purposes}
\label{sec:affine-accesses}
For various transformations, PolyBlocks relies on analysis of affine memory
access patterns. In this section, we show a simple example.

Consider the access to the 4-d tensor load in
Figure~\ref{fig:mod_div_load}. Does the access have contiguity
along loop IV `\%j'? Affine expressions, maps, and load/store operations in
MLIR's affine dialect make such analysis easy. The transcript has four
dimensions, and `\%j' appears in both the third and fourth dimensions. Most
low-level compilers' analyses are unable to reason about vectorizability here.
Although the penultimate subscript also changes with '\%j', the access does exhibit
spatial reuse in intervals of two, but not in intervals of four or eight. As a
result, it can be vectorized to use 2xf16 vector loads, which PolyBlocks'
vectorization pass performs given target information.  This is an example of
simple near O(n) time checks (in the IR expression lengths) that can resolve
contiguity analysis queries.

\begin{figure}[!htb]
\begin{lstlisting}[language=MLIR,basicstyle=\tiny\ttfamily]
affine.for %i = 0 to 32 {
  affine.for %j = 0 to 32 {
    %v = affine.load %memref[(%i + %ii * 32) floordiv 519684,
           (((%i + %ii * 32) mod 519684) floordiv 254) * 2 + (((%j + %jj * 32) floordiv 10) mod 16) floordiv 4,
           ((%i + %ii * 32) mod 254) * 2 + ((%j + %jj * 32) floordiv 10) mod 4,
           (%j + %jj * 32) mod 10
         ] : memref<8x4094x510x10xf16>
    affine.store %v, %13[%i, %j] : memref<32x40xf16, 3>
  }
}
\end{lstlisting}
\vskip -12pt
\caption{A specialized packing into an on-chip memory
  buffer: note contiguity in two-sized intervals along \%j.
\label{fig:mod_div_load}}
\vskip -12pt
\end{figure}

\subsection{Slicing-based affine fusion}
\label{sec:fusion}
The goal of this section is to first provide a high-level background of
slicing-based affine fusion, the basic infrastructure of which has been
available in the official MLIR repository. We do not claim that as a
contribution of this work. We describe how that infrastructure was customized
and extended to build a fusion pass in PolyBlocks with various heuristics and cost
modeling with extensibility for other targets.

Slicing-based fusion goes beyond traditional loop
fusion~\cite{wolfe95book,allenkennedybook} in the compiler literature. For a
given producer-store (the source), a consumer load (the destination) and a given
destination loop depth to fuse into, the approach computes the slice of the
producer that is to be executed to provide the dependent data to the consumer.
Consequently, the producer can be pulled in and materialized to compute the
required memref locations at that depth, leading to a shorter reuse distance, a
smaller temporary, or elimination of the temporary (register reuse in the case of
innermost fusion). Conversely, a consumer can be pulled into the producer with a
similar approach. Our fusion pass supports both the producer-into-consumer and
the consumer-into-producer fusion. Unlike traditional loop fusion, slicing-based
fusion can lead to the introduction of redundant computation, which has to be
carefully controlled through a cost model, for e.g., 0-10\% of redundant
computation can be tolerated.

\begin{figure}[htbp]
\begin{minipage}{0.52\linewidth}
\begin{lstlisting}[language=MLIR,basicstyle=\tiny\ttfamily]
  affine.for %i = 0 to 3 {
    affine.for %j = 0 to 4096 {
      affine.for %k = 0 to 4092 {
        affine.store %cst_3, %0[%i, %j, %k] : memref<3x4096x4092xf32>
      }
    }
  } {polyblocks.kind = "broadcast", polyblocks.target = "gpu"}
  affine.for %i = 0 to 3 {
    affine.for %j = 0 to 4096 {
      affine.for %k = 0 to 4092 {
        %7 = affine.load %arg0[%i, %j, %k] : memref<3x4096x4096xf32>
        %8 = affine.load %0[%i, %j, %k] : memref<3x4096x4092xf32>
        %9 = arith.mulf %7, %cst_1 : f32
        %10 = arith.addf %8, %9 : f32
        %11 = affine.load %arg0[%i, %j, %k + 1] : memref<3x4096x4096xf32>
        %12 = arith.mulf %11, %cst_0 : f32
        %13 = arith.addf %10, %12 : f32
        %14 = affine.load %arg0[%i, %j, %k + 2] : memref<3x4096x4096xf32>
        %15 = arith.mulf %14, %cst : f32
        %16 = arith.addf %13, %15 : f32
        affine.store %16, %0[%i, %j, %k] : memref<3x4096x4092xf32>
      }
    }
  } {polyblocks.kind = "stencil", polyblocks.target = "gpu"}
  %1 = memref.alloc() : memref<3x4092x4092xf32>
  affine.for %i = 0 to 3 {
    affine.for %j = 0 to 4092 {
      affine.for %k = 0 to 4092 {
        affine.store %cst_3, %1[%i, %j, %k] : memref<3x4092x4092xf32>
      }
    }
  } {polyblocks.kind = "broadcast", polyblocks.target = "gpu"}
  affine.for %i = 0 to 3 {
    affine.for %j = 0 to 33 {
      affine.for %k = 0 to 682 {
        affine.for %l = 0 to 1 {
          affine.for %m = 0 to 124 {
            affine.for %n = 0 to 6 {
              %7 = affine.load %0[%i + %l, %m + %j * 124,
                     %n + %k * 6] : memref<3x4096x4092xf32>
              %8 = affine.load %1[%i + %l, %m + %j * 124,
                     %n + %k * 6] : memref<3x4092x4092xf32>
              %9 = arith.mulf %7, %cst_1 : f32
              %10 = arith.addf %8, %9 : f32
              %11 = affine.load %0[%i + %l, %m + %j * 124 + 1,
                     %n + %k * 6] : memref<3x4096x4092xf32>
              %12 = arith.mulf %11, %cst_0 : f32
              %13 = arith.addf %10, %12 : f32
              %14 = affine.load %0[%i + %l, %m + %j * 124 + 2,
                     %n + %k * 6] : memref<3x4096x4092xf32>
              %15 = arith.mulf %14, %cst : f32
              %16 = arith.addf %13, %15 : f32
              affine.store %16, %1[%i + %l, %m + %j * 124,
                     %n + %k * 6] : memref<3x4092x4092xf32>
            }
          }
        }
      }
    }
  } {polyblocks.kind = "stencil", polyblocks.target = "gpu"}
\end{lstlisting}
\subcaption{Original code.}
\end{minipage}
\begin{minipage}{0.42\linewidth}
\begin{lstlisting}[language=MLIR,basicstyle=\tiny\ttfamily]
affine.for %i = 0 to 3 {
	affine.for %j = 0 to 33 {
		affine.for %k = 0 to 682 {
			%alloc_7 = memref.alloc() : memref<1x126x6xf32, 3>
			affine.for %l = affine_map<(d0) -> (d0 * 124)>(%j)
          to affine_map<(d0) -> (d0 * 124 + 126)>(%j) {
				affine.for %m = affine_map<(d0) -> (d0 * 6)>(%k)
            to affine_map<(d0) -> (d0 * 6 + 6)>(%k) {
					affine.store %cst_3, %alloc_7[0, %j * -124 + %l,
              %k * -6 + %m] : memref<1x126x6xf32, 3>
					%0 = affine.load %arg0[%i, %l, %m]
                  : memref<3x4096x4096xf32>
					%1 = arith.mulf %0, %cst_1 : f32
					%2 = arith.addf %1, %cst_3 : f32
					%3 = affine.load %arg0[%i, %l, %m + 1]
                  : memref<3x4096x4096xf32>
					%4 = arith.mulf %3, %cst_0 : f32
					%5 = arith.addf %2, %4 : f32
					%6 = affine.load %arg0[%i, %l, %m + 2]
                  : memref<3x4096x4096xf32>
					%7 = arith.mulf %6, %cst : f32
					%8 = arith.addf %5, %7 : f32
					affine.store %8, %alloc_7[0, %j * -124 + %l,
               %k * -6 + %m] : memref<1x126x6xf32, 3>
				}
			}
			affine.for %l = 0 to 1 {
				affine.for %m = 0 to 124 {
					affine.for %n = 0 to 6 {
						affine.store %cst_3, %alloc[%i + %l, %j * 124 + %m,
                %k * 6 + %n] : memref<3x4092x4092xf32>
						%0 = affine.load %alloc_7[%l, %m, %n]
                    : memref<1x126x6xf32, 3>
						%1 = arith.mulf %0, %cst_1 : f32
						%2 = arith.addf %1, %cst_3 : f32
						%3 = affine.load %alloc_7[%l, %m + 1, %n]
                    : memref<1x126x6xf32, 3>
						%4 = arith.mulf %3, %cst_0 : f32
						%5 = arith.addf %2, %4 : f32
						%6 = affine.load %alloc_7[%l, %m + 2, %n]
                    : memref<1x126x6xf32, 3>
						%7 = arith.mulf %6, %cst : f32
						%8 = arith.addf %5, %7 : f32
						affine.store %8, %alloc[%i + %l, %m + %j * 124,
                %n + %k * 6] : memref<3x4092x4092xf32>
					}
				}
			}
			memref.dealloc %alloc_7 : memref<1x126x6xf32, 3>
		}
	}
} {polyblocks.kind = "stencil", polyblocks.target = "gpu"}
\end{lstlisting}
\subcaption{Fused code.}
\end{minipage}
\vskip -6pt
\caption{Affine slicing-based fusion. The first stencil nest is
pulled into the second one by computing the producer slice needed by the
consumer. Fusion happens into right under the 682-iteration \%k loop of
Figure~\ref{fig:affine-fusion}(a), i.e., given an \%i, \%j, \%k, the slice of
the producer needed.\label{fig:affine-fusion}}
\end{figure}

To determine the validity of such slicing-based fusion precisely, various checks
need to be performed, including the presence of intervening conflicting memory
operations and whether the source can be safely erased post-fusion. In a very
small set of cases, we end up using the integer set library~\cite{isl,isl-web}
when simpler techniques are insufficient.  This usage can be replaced with the
FPL library~\cite{grosser20oopsla,pitchanathan21oopsla} already available in
MLIR when the latter has support on par with ISL. To compute the slice itself,
affine memory dependence constraints are used to express the slice in terms of
the destination loop iterators.

While evaluating a specific fusion, we consider criteria such as preservation of
parallelism, preservation of vectorizability, the amount of redundant
computation added, and whether the fusion will eventually lead to the
elimination of intermediate buffers, i.e., the creation of a private memref or a
scalar in the case of register-level (typically innermost loop) fusion, and
whether the size of private memref fits in available fast chip memory. We also
make use of CBrick kinds (Section~\ref{sec:polyblocks-kinds}) to speed up the
checks and eliminate undesirable fusion with a simple set of rules. For example,
the fusion between two matmul or two reduction cbrick kinds is not attempted,
except separately in scenarios covered in Section~\ref{sec:fused-attention}.
Similarly, the fusion of any non-trivial operation into broadcasting consumers is
disallowed.

In Figure~\ref{fig:affine-fusion}, a fusion is performed between two
stencil-like operations that leads to an imperfect nest and the creation of
a 3~KB buffer (\%alloc\_7) in the local on-chip memory, completely eliminating
a trip to global memory and a global memory allocation. Both XLA and Inductor
are unable to perform such fusion across such operators. This specific fusion
is easily performed as a result of modeling on affine nests (potentially
imperfectly nested) and affine memory operations.  The stencils here were
lowered forms from depth-wise convolution operators expressed in PyTorch. For
GPUs, IR such as this leads to a single fused kernel.

\subsection{Multi-level tiling and fusion approach}
While it is well-known how to perform tiling on loop nests, its interplay with
fusion is complex. The end-result one desires is a nest that is tiled and fused.
However, neither fusing and then tiling, nor tiling and then fusing, is
sufficient in general and in many important cases; both could lead to
undesirable results. Consider the situation of slicing-based fusion. The slice
that is being pulled in gets effectively tiled as per its destination. The
tiling expressed on the source nest, if any, would be lost. It is thus
sufficient to tile the destination appropriately and then pull in untiled
sources (producers or consumers).

{\bf Two-phase approach:} Tiling a nest that will be the destination of
subsequent fusions, either of its producers or of its consumers, is thus the
first key step.  Once all fusions determined to be profitable into such tiled
nests are performed, the remaining nests can be tiled either for locality or
parallelism purposes. We follow such a two-round approach. We first tile a set
of key nests, which typically include all matmuls, convolutions, and destination
stencils. We then perform the slicing-based fusion described in
Section~\ref{sec:fusion} on all nests. In the second round, all remaining
untiled nests are tiled. With this approach, the fusion of pointwise nests with
other pointwise nests happens in their untiled form, which is simple and
straightforward. Fusion into matmuls and convolutions happens into tiled nests.

Fusion into matmuls and convolutions has to happen after the generation of fast
buffers, i.e., global to on-chip memory movement code generation.  As such, fast
buffer generation for matmuls and convolutions is performed early.  Fusion
subsequently happens into the copy-in and copy-out nests of such tiled nests,
which is the desired outcome. Data movement code generation for the remaining
nests where accesses exhibit temporal reuse is performed late in the pipeline.

Each level of parallelism represented by a contiguous band of {\it affine.for}
loops is turned into a multi-dimensional {\it affine.parallel} operation after all
desired optimization on `affine.for` nests (cf.
Figure~\ref{fig:mma-mapped-nest}). The interactions between tiling and fusion
also require careful consideration in the selection of tile sizes and the choice
to generate fast buffers (whether to use on-chip memory for a specific
memref/access) since the eventual on-chip memory utilization changes after
fusion has been performed. Our pipeline is able to spill buffers from fast
memory to global memory much later in its stages if the utilization goes beyond
the on-chip memory capacity, something that may rarely happen due to the
inability to model interactions between passes that are distant in the pipeline.

\subsection{Fusion and tiling of the attention layer}
\label{sec:fused-attention}

In this section, we describe how we automatically achieve the fusion of all
operations in the attention layer~\cite{vaswani17nips} through a composition of
simpler loop and code motion transformations. The fusion required here to achieve
state-of-the-art performance~\cite{dao20nips} is normally outside the scope of
both standard compiler fusion~\cite{wolfe95book,allenkennedybook}, slicing-based
fusion techniques~\cite{halide,xla,zhao24tocs} , including those described in
Section~\ref{sec:fusion} or any other techniques that rely on data dependences.
Note that such a fusion can be realized only because it is possible to compute the
softmax part online~\cite{online-softmax-milakov18arxiv}. PolyBlocks primarily
employs two optimization passes for this purpose, {\it reduce-reduce-fusion} and
{\it wmma-fusion}, to minimize expensive memory accesses in an attention layer
computation. These passes target DRAM and on-chip memory accesses, respectively,
specifically addressing the large output of $Q*K^T$.

The reduce-reduce-fusion pass operates on the attention layer, which is
initially represented as a sequence of affine nests. The pass's objective is to
fuse these nests into a single affine nest. It identifies a series of nests for
fusion based on heuristics that consider factors such as matrix multiplication
problem sizes, the number of intervening reduction nests and their
interdependencies. Memory dependences among identified nests form a
directed acyclic graph (DAG). The reduce-reduce-fusion pass fuses them into the
sink node in reverse topological order. A notable challenge is in supporting
the fusion of a producer reduction nest into a consumer reduction nest within a
consumer reducing loop.  This is crucial to achieve good performance for the
attention layer. However, such a fusion violates memref dependences on the
result memref of the producer reduction. To overcome this, the fusion is
performed despite the dependence, and additional operations are then introduced
to preserve the original program semantics. Our approach automatically extracts
the necessary correction computation from the fused nest.  This correction is
then inserted right before the consumer reduction load, ensuring that the consumer
reduction result remains consistent with the last executed iteration of the
reduction loop.

In the output of the reduce-reduce-fusion pass, a consumer reduction result is
corrected whenever a producer reduction result is updated. However, this
correction should only occur once per tile of the producer reduction result to
minimize computational overhead. To achieve this, we employ a second pass called
{\it attention-matmul-outlining}. This pass transforms a tiled fused-attention
nest, outlining the matrix multiplications and separating the correction
computation from the reduction.

Our reduce-reduce-fusion pass can handle the Attention variations supported by
FlexAttention~\cite{dong24arxiv} with little to no extra effort, because many
of the variations supported by FlexAttention, when lowered to affine nests,
result in one or more pointwise nests operating on the output of $Q*K^T$,
and generating an input for the subsequent softmax part, and the fusion of which
is already supported by our fusion utilities.

\textbf{wmma-fusion:} While the reduce-reduce-fusion pass eliminates DRAM
round-trips for the output of the {\it Query} and {\it Value} matrix
multiplication, achieving peak performance also necessitates avoiding
round-trips to on-chip memory.  This is where a third pass comes into play, the
wmma-fusion pass. This pass works on a fused attention nest that comprises the
two matrix multiplications, along with other nests required for softmax
computation and the ``correction'' operations. At this stage, all nests are
fetching data from on-chip memory, and the matrix multiplication nests have
already been mapped to matrix units if any (Warp Matrix Multiply-Accumulate
operations in the case of NVIDIA/AMD GPUs).  The wmma-fusion pass first maps all
non-matmul nests within the fused attention to WMMA/matrix operations, ensuring
compatibility with the existing matmul mappings.  This mapping is facilitated by
our matrix-mapping utilities. Once all nests are mapped, they are fused into the
first matrix multiplication using existing fusion utilities
(Section~\ref{sec:fusion}). The fused nest is then fully unrolled, and scalar
replacement is performed, effectively eliminating round-trips to on-chip shared
memory, i.e., results are fully reused within registers.

\subsection{Transforming and mapping convolutions with on-the-fly packing}
\label{sec:conv-mapping}

It is well-known that convolutions can be computed using matrix multiplication
operations. The left-hand side (LHS) and right-hand side (RHS) of the matrix
multiplication are constructed using the input image and filter weights,
respectively. Each row of the LHS matrix is formed by flattening the image
elements overlapped by the kernel window, and each column of the RHS matrix is
formed by flattening the elements in the kernel window. Explicit GEMM and
implicit GEMM are two terms used to specify whether such a reorganization is
performed globally on the data or on-the-fly on tiles.

Materializing the LHS matrix can cause significant memory overhead due to
overlaps in the sliding window. Employing on-the-fly packing as shown in
Figure~\ref{fig:conv-packing} addresses this issue, where we form a tile of the
LHS matrix in the fast memory by loading the required elements from the original
image tensor. Similarly, a tile of the RHS matrix is formed using the filter
weights. Later, matrix multiplication is performed on the LHS and RHS tiles, and
the elements of the output matrix are stored back to the original output tensor.
We perform such a transformation by first generating a 3-d loop form of the
matrix multiplication nest (Figure~\ref{fig:conv-as-matmul}) that operates on
the original tensors, but with the right affine accesses. Later, this nest is
tiled and fast buffers are generated to realize on-the-fly packing shown in
Figure~\ref{fig:conv-packing}.

Partial tiles are formed if the dimension of the resultant matrices is not a
multiple of the chosen tile size or the hardware-supported matrix unit. To
ensure correct results, we pad the partial contracting dimension of the LHS tile
with zeros in the fast memory (Figure~\ref{fig:conv-packing}).

Note that no new operations were required in MLIR to realize this
transformation. Also, this transformation works for all 2-d convolutions with
different configurations (stride, dilation, and padding kinds) and transpose
convolutions as well.  The evaluation presented in Section~\ref{sec:evaluation}
exercises all of this.

A related approach is the implicit GEMM algorithm included with
Cutlass~\cite{cutlass-implicit-gemm}.  However, to the best of our knowledge,
none of the existing high-level compilers are able to auto-generate such a
packing and mapping from loop nests for the entire diversity of convolutions.

\begin{figure}[htbp]
\begin{minipage}{0.42\linewidth}
\begin{lstlisting}[language=MLIR,basicstyle=\fourpointsize\ttfamily]
affine.for %i = 0 to 3528 {
  affine.for %j = 0 to 512 {
    affine.for %k = 0 to 1143 {
      %0 = affine.load %image_memref[%i floordiv 441,
             ((%i mod 441) floordiv 21) * 2 + (%k floordiv 127) floordiv 3,
             %i * 2 - (%i floordiv 21) * 42 + %k floordiv 127
             - ((%k floordiv 127) floordiv 3) * 3,
             %k mod 127] : memref<8x44x44x127xf16>
      %1 = affine.load %output_memref[%i floordiv 441, (%i mod 441) floordiv 21,
                                      %i mod 21, %j] : memref<8x21x21x512xf32>
      %2 = affine.load %weights_memref[(%k floordiv 127) floordiv 3,
             (%k floordiv 127) mod 3, %k mod 127, %j] : memref<3x3x127x512xf16>
      %3 = arith.extf %0 : f16 to f32
      %4 = arith.extf %2 : f16 to f32
      %5 = arith.mulf %3, %4 : f32
      %6 = arith.addf %5, %1 : f32
      affine.store %6, %output_memref[%i floordiv 441, (%i mod 441) floordiv 21,
                                      %i mod 21, %j] : memref<8x21x21x512xf32>
    }
  } {polyblocks.parallel}
} {polyblocks.kind = "matmul", polyblocks.matmul_info = {virtual_matmul}, polyblocks.parallel}
\end{lstlisting}
\subcaption{Convolution lowered as 3-d loop matrix multiplication nest using affine accesses.\label{fig:conv-as-matmul}}
\end{minipage}
\begin{minipage}{0.55\linewidth}
\begin{lstlisting}[language=MLIR,basicstyle=\fourpointsize\ttfamily]
affine.for %i = 0 to 3528 step 128 {
  affine.for %j = 0 to 512 step 128 {
    %input_fast_buf = memref.alloc() : memref<128x32xf16, 3>
    %weights_fast_buf = memref.alloc() : memref<32x128xf16, 3>
    %output_fast_buf = memref.alloc() : memref<128x128xf32, 3>
    // Output packing nest.
    affine.for %ii = 0 to min affine_map<(d0) -> (128, -d0 + 3528)>(%i) {
      affine.for %jj = 0 to 128 {
        %0 = affine.load %output_memref[(%i + %ii) floordiv 441, ((%i + %ii) mod 441) floordiv 21,
                                            (%i + %ii) mod 21, %j + %jj] : memref<8x21x21x512xf32>
        affine.store %0, %output_fast_buf[%ii, %jj] : memref<128x128xf32, 3>
      } {polyblocks.parallel}
    } {copy_nest, polyblocks.parallel}
    affine.for %k = 0 to 1143 step 32 {
      // Input packing nest.
      affine.for %ii = 0 to 128 {
        affine.for %kk = 0 to 32 {
          %0 = affine.load %image_memref[((%i + %ii) floordiv 441) mod 8,
                   (((%i + %ii) mod 441) floordiv 21) * 2 + ((%k + %kk) floordiv 127) floordiv 3,
									 (%i * 2 + %ii * 2 - ((%i + %ii) floordiv 21) * 42 + (%k + %kk) floordiv 127
                   - (((%k + %kk) floordiv 127) floordiv 3) * 3) mod 44,
                   (%k + %kk) mod 127] : memref<8x44x44x127xf16>
          affine.store %0, %input_fast_buf[%ii, %kk] : memref<128x32xf16, 3>
        } {polyblocks.parallel}
      } {copy_nest, polyblocks.parallel}
      // Weights packing nest.
      affine.for %kk = 0 to 32 {
        affine.for %jj = 0 to 128 {
          %0 = affine.load %weights_memref[(((%k + %kk) floordiv 127) floordiv 3) mod 3,
                 ((%k + %kk) floordiv 127) mod 3,
                 (%k + %kk) mod 127, %j + %jj] : memref<3x3x127x512xf16>
          affine.store %0, %weights_fast_buf[%kk, %jj] : memref<32x128xf16, 3>
        } {polyblocks.parallel}
      } {copy_nest, polyblocks.parallel}
      // Zero pad the partial tile in the fast memory.
      affine.for %kk = 1143 to affine_map<(d0) -> (d0 + 32)>(%k) {
        affine.for %jj = 0 to 128 {
          affine.store %zero, %weights_fast_buf[%kk - %k, %jj] : memref<32x128xf16, 3>
        }
      }
      // Matrix multiplication compute nest.
      affine.for %ii = 0 to 128 step 64 {
        affine.for %jj = 0 to 128 step 64 {
          affine.for %kk = 0 to 32 step 16 {
            affine.for %iii = 0 to 64 {
              affine.for %jjj = 0 to 64 {
                affine.for %kkk = 0 to 16 {
                  %0 = affine.load %input_fast_buf[%ii + %iii, %kk + %kkk] : memref<128x32xf16, 3>
                  %1 = affine.load %weights_fast_buf[%kk + %kkk, %jj + %jjj] : memref<32x128xf16, 3>
                  %2 = affine.load %output_fast_buf[%ii + %iii, %jj + %jjj] : memref<128x128xf32, 3>
                  %3 = arith.extf %0 : f16 to f32
                  %4 = arith.extf %1 : f16 to f32
                  %5 = arith.mulf %3, %4 : f32
                  %6 = arith.addf %5, %2 : f32
                  affine.store %6, %output_fast_buf[%ii + %iii, %jj + %jjj] : memref<128x128xf32, 3>
                }
              }
            }
          } {polyblocks.tile_space}
        } {polyblocks.tile_space}
      } {polyblocks.tile_space}
    } {compute_nest, polyblocks.tile_space}
    // Output unpacking nest.
    affine.for %ii = 0 to min affine_map<(d0) -> (128, -d0 + 3528)>(%i) {
      affine.for %jj = 0 to 128 {
        %0 = affine.load %output_fast_buf[%ii, %jj] : memref<128x128xf32, 3>
        affine.store %0, %output_memref[(%i + %ii) floordiv 441, ((%i + %ii) mod 441) floordiv 21,
                                        (%i + %ii) mod 21, %j + %jj] : memref<8x21x21x512xf32>
      } {polyblocks.parallel}
    } {copy_nest, polyblocks.parallel}
  } {polyblocks.parallel, polyblocks.tile_space}
} {polyblocks.kind = "matmul"}
\end{lstlisting}
\subcaption{On-the-fly packing for convolution.\label{fig:conv-packing}}
\end{minipage}
\vskip -6pt
\caption{On-the-fly tiled mapping for convolutions.}
\end{figure}

\subsection{Mapping to matrix units}
Several AI chips include specialized hardware to perform matrix-matrix
multiplication. PolyBlocks includes a pass that performs the necessary analysis and
transformation of affine nests to map from load, store, and compute on
elemental-typed memrefs to those of 2-d vector types, which can then be
converted to target-specific types. In the case of GPUs, the operations are
converted to those on memrefs of subgroup MMA type of size 16x16. We introduced
a memref view-like op {\it polyblocks.matrix\_cast} to obtain a view of the memref
on 2-d vector types; this view-like op is deabstracted and lowered subsequently soon.
As a result of implementing matricization on affine loop nests, we avoid the
need to provide a mechanism to implement such a mapping on various kinds of
nests like matmul, convolutions, matmuls within a fused attention nest (as related
to Section~\ref{sec:fused-attention}), and tensor contractions ensuing
from einsum computation lowering. A nest mapped to GPU tensor cores is shown in
Figure~\ref{fig:mma-mapped-nest}.

\begin{figure}[htbp]
\begin{lstlisting}[language=MLIR,basicstyle=\fourpointsize\ttfamily]
// Grid of thread blocks.
affine.parallel (%arg3) = (0) to (379) {
  %alloc = memref.alloc() : memref<128x136xf16, 3>
  %alloc_5 = memref.alloc() {alignment = 16 : i64} : memref<128x16xf16, 3>
  %7 = memref.vector_cast %alloc_5 : memref<128x16xf16, 3> to memref<128x2xvector<8xf16>, 3>
  %alloc_6 = memref.alloc() : memref<128x12xf32, 3>
  affine.parallel (%arg4, %arg5) = (0, 0) to (min(128, %arg3 * -128 + 48400), 3) {
    affine.store %cst_0, %alloc_6[%arg4, %arg5] : memref<128x12xf32, 3>
  }
  // Warp-level parallelism.
  affine.parallel (%arg4) = (0) to (4) {
    %8 = affine.apply affine_map<(d0) -> (d0 * 32)>(%arg4)
    %9 = gpu.subgroup_mma_load_matrix %alloc_6[%8, %c0] {leadDimension = 12 : index} : memref<128x12xf32, 3> -> !gpu.mma_matrix<32x8xf32, "COp">
    // Threads of a thread block. Copy-in from global memory.
    affine.parallel (%arg5, %arg6) = (0, 0) to (128, 128) {
      %11 = affine.load %memref[0, ((%arg5 + %arg3 * 128) floordiv 220 + (%arg6 floordiv 3) floordiv 5) mod 224,
        ((%arg5 + %arg3 * 128) mod 220 + %arg6 floordiv 3 - ((%arg6 floordiv 3) floordiv 5) * 5) mod 224, %arg6 mod 3] : memref<1x224x224x3xf16>
      affine.store %11, %alloc[%arg5, %arg6] : memref<128x136xf16, 3>
    }
    // Threads of a thread block. Copy-in from global memory.
    affine.parallel (%arg5, %arg6) = (0, 0) to (128, 8) {
      %11 = affine.load %memref_1[((%arg5 floordiv 3) floordiv 5) mod 5, (%arg5 floordiv 3) mod 5, %arg5 mod 3, %arg6 mod 3] : memref<5x5x3x3xf16>
      affine.store %11, %alloc_5[%arg5, %arg6] : memref<128x16xf16, 3>
    }
    // Zero init.
    affine.parallel (%arg5) = (0) to (53) {
      affine.store %cst, %7[%arg5 + 75, 0] : memref<128x2xvector<8xf16>, 3>
    }
    %10 = affine.for %arg5 = 0 to 8 iter_args(%arg6 = %9) -> (!gpu.mma_matrix<32x8xf32, "COp">) {
      %11 = affine.apply affine_map<(d0) -> (d0 * 16)>(%arg5)
      %12 = affine.apply affine_map<(d0) -> (d0 * 32)>(%arg4)
      %13 = gpu.subgroup_mma_load_matrix %alloc[%12, %11] {leadDimension = 136 : index} : memref<128x136xf16, 3> -> !gpu.mma_matrix<32x16xf16, "AOp">
      %14 = gpu.subgroup_mma_load_matrix %alloc_5[%11, %c0] {leadDimension = 16 : index} : memref<128x16xf16, 3> -> !gpu.mma_matrix<16x8xf16, "BOp">
      // Warp-level primitives.
      %15 = gpu.subgroup_mma_compute %13, %14, %arg6 : !gpu.mma_matrix<32x16xf16, "AOp">, !gpu.mma_matrix<16x8xf16, "BOp"> -> !gpu.mma_matrix<32x8xf32, "COp">
      affine.yield %15 : !gpu.mma_matrix<32x8xf32, "COp">
    }
    gpu.subgroup_mma_store_matrix %10, %alloc_6[%8, %c0] {leadDimension = 12 : index} : !gpu.mma_matrix<32x8xf32, "COp">, memref<128x12xf32, 3>
  }
  memref.dealloc %alloc_5 : memref<128x16xf16, 3>
  memref.dealloc %alloc : memref<128x136xf16, 3>
  // Threads of a thread block. Copy-out to global memory.
  affine.parallel (%arg4, %arg5) = (0, 0) to (min(128, %arg3 * -128 + 48400), 3) {
    %8 = affine.load %alloc_6[%arg4, %arg5] : memref<128x12xf32, 3>
    %9 = arith.truncf %8 : f32 to f16
    affine.store %9, %memref_3[0, (%arg4 + %arg3 * 128) floordiv 220, (%arg4 + %arg3 * 128) mod 220, %arg5] : memref<1x220x220x3xf16>
  }
  memref.dealloc %alloc_6 : memref<128x12xf32, 3>
}
\end{lstlisting}
\vskip -12pt
\caption{IR showing an optimized nest mapped to GPU tensor cores with multi-level parallelism encoded.\label{fig:mma-mapped-nest}}
\end{figure}

\subsection{Other optimizations}
Appendices \ref{sec:quantization} and \ref{sec:compute-copy-overlap}, part of the
supplemental material, describe two more optimizations: (1) quantization support
and how all optimizations described apply out of the box to it, and (2) overlap
of compute with data copy.  Other standard transformations like register tiling,
generation of data movement code, vectorization, and unrolling are part of our
pipeline, and we do not describe these in more detail. All of these passes rely
on analysis utilities that determine the amount of temporal and spatial reuse,
access pattern analysis (Section~\ref{sec:affine-accesses}), parallelism
estimation, and polyblocks kind information (Appendix~\ref{sec:polyblocks-kinds}).  We
perform all of these transformations on affine loop nests in S3.

\section{Experimental Evaluation}
\label{sec:evaluation}
The evaluation was carried out on an NVIDIA A100 GPU and, in some cases, on an
NVIDIA A10 GPU (for batch size 1). As the baselines of Torch Inductor or
JAX-XLA rely on hand-tuned highly-optimized vendor libraries, our expectation
is not to be always better than these baselines, but to be competitive. The
libraries from vendors always get a head start in terms of reaching peak
performance of hardware and are often optimized for the sizes found in popular
models. Thus, getting close to the performance of library-based approaches
still demonstrates the power of fully code-generating compilers, and that they
are promising for future hardware not well-served by the manually-developed
kernel ecosystem. Apart from the presented evaluation, PolyBlocks delivers the
same capabilities and optimizations for TensorFlow, and for CPUs and AMD GPUs.
All PolyBlocks compiled code evaluated here is fully automatically generated with
no manual intervention or auto-tuning, and with same unmodified PyTorch and JAX
as input as with eager executors or Inductor/XLA. Compilation times and PyTorch
model coverage with PolyBlocks is provided in
Appendix~\ref{sec:compilation-time-coverage}.

\subsection{Experimental setup}
For all benchmarking, we discard a certain number of warmup runs and take the
average of the subsequent 10 or 100 runs, depending on execution time lengths.
All execution times reported are the  sum total NVIDIA Nsight Systems (nsys) kernel
execution times.  We lock the clocks for the A10 and A100 GPUs to the
recommended 1695~MHz and 1410~MHz, respectively, uniformly for all
implementations. The CUDA driver version used was 570.86.15 (CUDA 12.8).

All models are inference workloads that are run in the mixed precision mode,
i.e., f16 input and fp32 output for all tensor core targeted computation.
Torch-eager and Inductor were used with automatic mixed precision (AMP).

\subsection{Performance comparison with Torch Inductor and Eager}

\begin{figure}[htb]
\includegraphics[width=\linewidth]{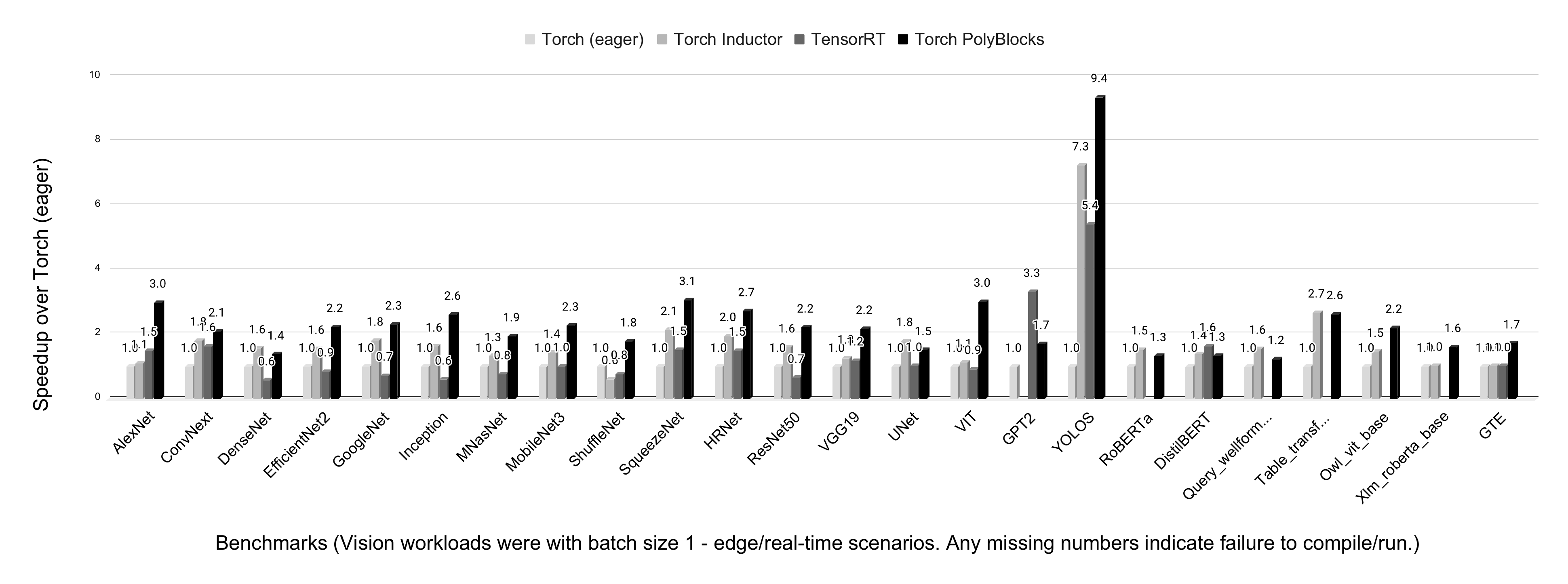}
\vskip -12pt
  \caption{
  PyTorch performance with various backends: eager, Inductor, TensorRT, and PolyBlocks. Batch size 1 on the NVIDIA A10.
  Torch 2.5.1, torch-tensorrt-2.5.0, TensorRT 10.3.0.
\label{fig:summary-bs-1}}
\vskip -12pt
\end{figure}
\begin{figure}[htb]
\includegraphics[width=\linewidth]{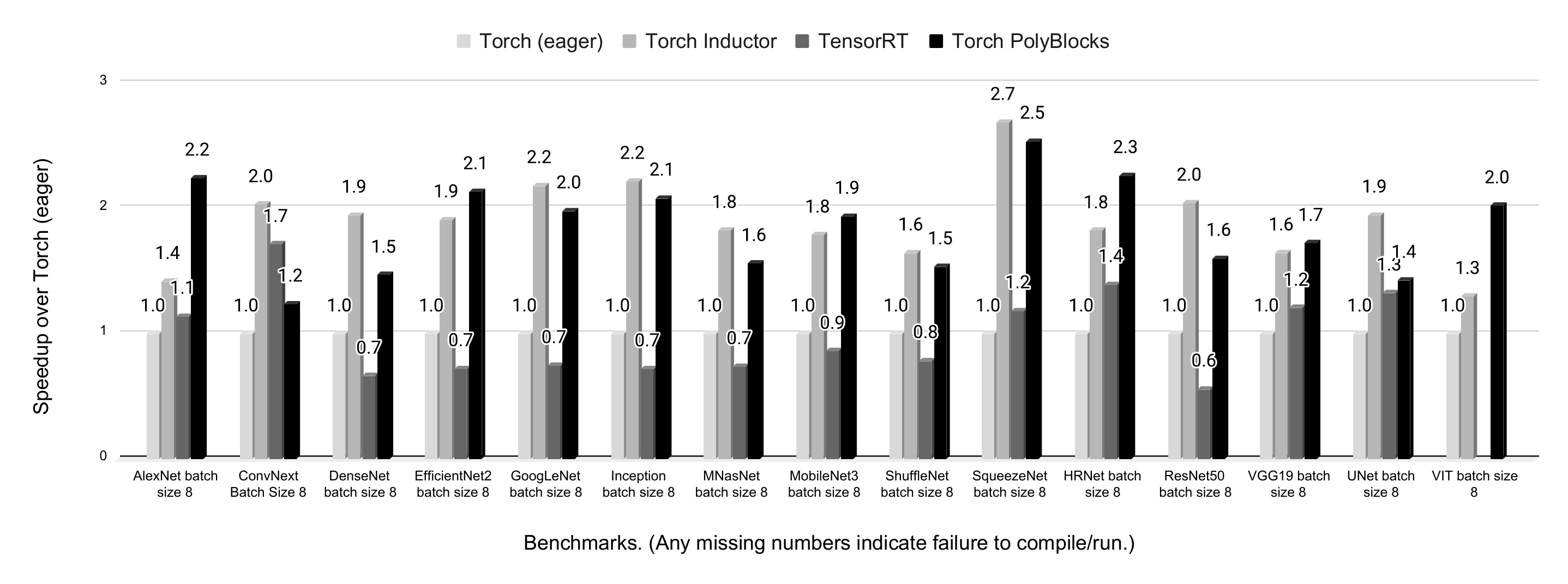}
\vskip -12pt
  \caption{
  PyTorch performance with various backends: eager, Inductor, TensorRT, and PolyBlocks. Batch size 8 on the NVIDIA A100.
  Torch 2.5.1, torch-tensorrt-2.5.0, TensorRT 10.3.0.
\label{fig:summary-bs-8}}
\end{figure}

We present our evaluation on PyTorch models from
Torchvision~\cite{TorchVision_maintainers_and_contributors_TorchVision_PyTorch_s_Computer_2016}
and Huggingface~\cite{huggingface}, selected to represent sufficient diversity
as far as being based on convolutional networks and transformers, vision and
NLP.  Figures~\ref{fig:summary-bs-1} and \ref{fig:summary-bs-8} show the results
on PyTorch models for batch sizes 1 and 8, respectively. While batch size 1
results, reflective of real-time inference, are on the A10 GPU, we use the A100
for batch size 8 which would lead to large utilization and thus more appropriate
for a larger GPU.

For batch size one, PyTorch with PolyBlocks as backend on these models is 2.15x as
fast on average (geomean) than eager execution, and 1.4x as fast as Inductor and
2.4x as fast as TensorRT. The benefits come from more or better fusion as
well as better code generated in some cases for individual operators as will be
shown in Section~\ref{sec:conv-matmul}.

For batch size 8, PolyBlocks is once again 1.8x as fast as eager on average
(geomean), and on par (0.97x) with Inductor. In general, we observe that
improvements over Inductor are higher with lower batch sizes. This is primarily
because the underlying libraries used by Torch Inductor are optimized for larger
batch sizes.

\subsection{JAX with PolyBlocks}
For JAX we
use Scenic models~\cite{dehghani2021scenic} and we use JAX-mixed-precision
(JMP)~\cite{jax-jmp} to specify a similar mixed precision mode. The mode is
specified by setting a JMP policy as \\ \texttt{\small policy =
jmp.get\_policy("float16")} and subsequently casting the model weights to f16
using \texttt{\small policy.cast\_to\_params(model\_weights)}. We observed that
this along with the f16 inputs was sufficient to trigger mixed-precision kernels
for JAX-XLA and JAX eager execution.

The PolyBlocks compiler engine with the same pass pipelines is able to compile JAX
with a single-line annotation like XLA does. We include a limited set of models
with JAX. Table~\ref{tab:polyblocks-jax} shows that JAX with PolyBlocks is 2.12x
(geomean) as fast as JAX eager execution and 1.15x as fast as XLA, an
established compiler backend for JAX.
\begin{table}[!htb]
  \vskip -5pt
	\caption{JAX with XLA vs PolyBlocks. On NVIDIA
	A100, JAX 5.0 (Jan 2025) with CUDA 12.9, Flax 0.10.4.\label{tab:polyblocks-jax}}
\vskip -6pt
\footnotesize
	\begin{tabularx}{0.9\linewidth}{l>{\centering\arraybackslash}p{0.1\linewidth}>{\centering\arraybackslash}p{0.15\linewidth}>{\centering\arraybackslash}p{0.15\linewidth}>{\centering\arraybackslash}p{0.12\linewidth}>{\centering\arraybackslash}p{0.12\linewidth}}
		\toprule \multirow{2}{*}{\textbf{Workload}}
    & \multicolumn{3}{c}{\bf JAX execution time (ms)}
		& \multicolumn{2}{c}{\bf Speedup with PolyBlocks} \\
		& \multirow{1}{*}{\textbf{
    eager}} & \multirow{1}{*}{\textbf{XLA}} & \multirow{1}{*}{\textbf{
    PolyBlocks}} & over eager & over XLA \\
		\midrule
		Resnet Batch 8	& 6.3	& 2.97	& 1.87	& 3.37	& 1.59 \\ Unet Batch 8	& 84.6
		& 60.1	& 69.1	& 1.22	& 0.87 \\
		Vit Batch 8	& 10.6	& 5.09	& 4.59 & 2.32	& 1.11 \\
		\bottomrule
	\end{tabularx}
  \vskip -10pt
\end{table}

\subsection{Individual operator performance: convolutions and matmuls}
\label{sec:conv-matmul}
While cross-operator fusion is the key strength of a compiler,
compiler-generated code will not perform well if its generated code is not
optimized for individual operators, especially for the compute-intensive ones
that provide significant opportunity to improve arithmetic intensity. We
typically expect compiler-generated code to be within 1.5x to 2x of the fastest
libraries on individual operators so that the benefits from fusion are
noticeable.

\begin{figure}[htbp]
  \subfloat[A10, batch size 1.]{
    \includegraphics[width=0.48\linewidth]{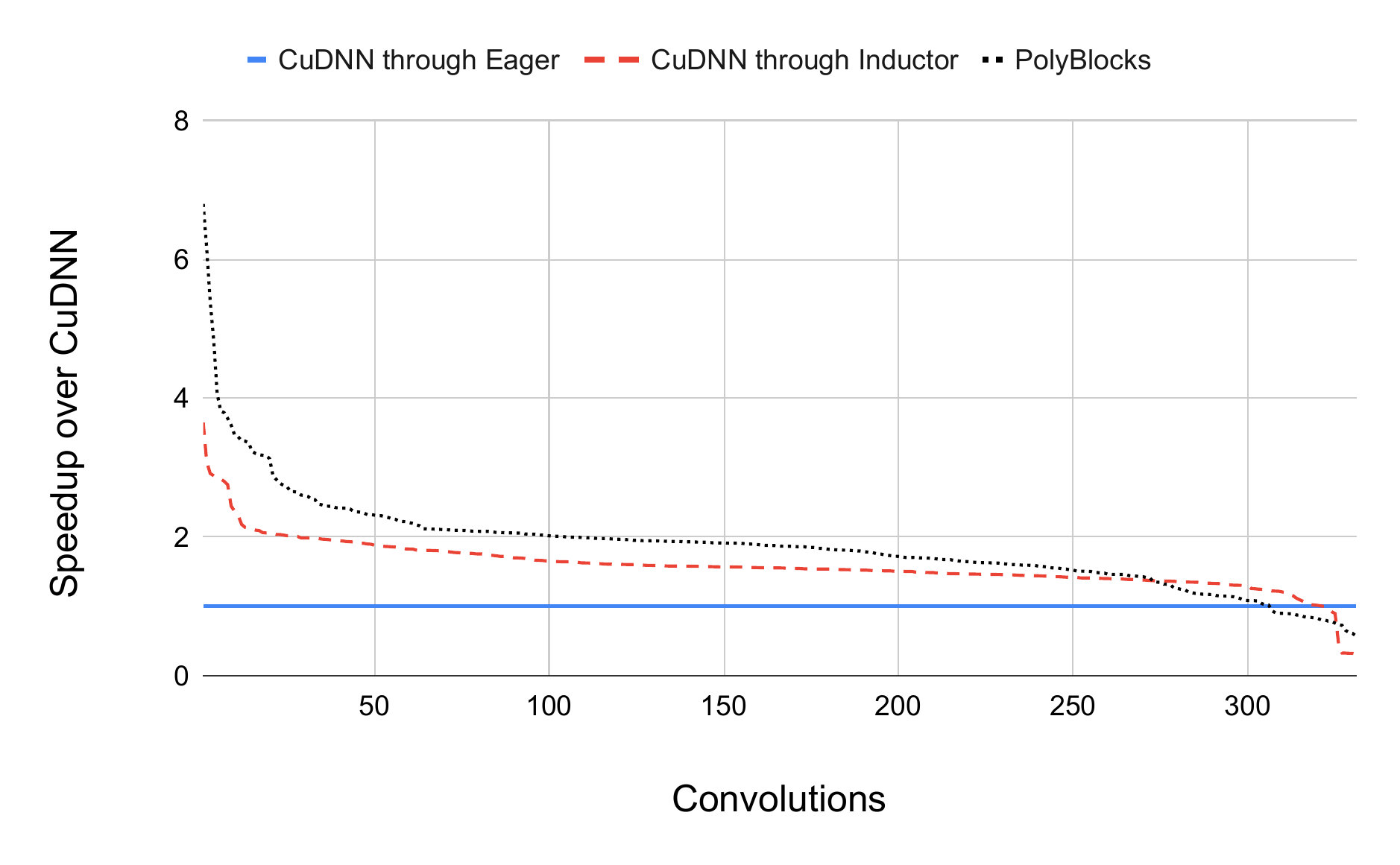} }
  \hfill \subfloat[A100, batch size 8.]{
    \includegraphics[width=0.48\linewidth]{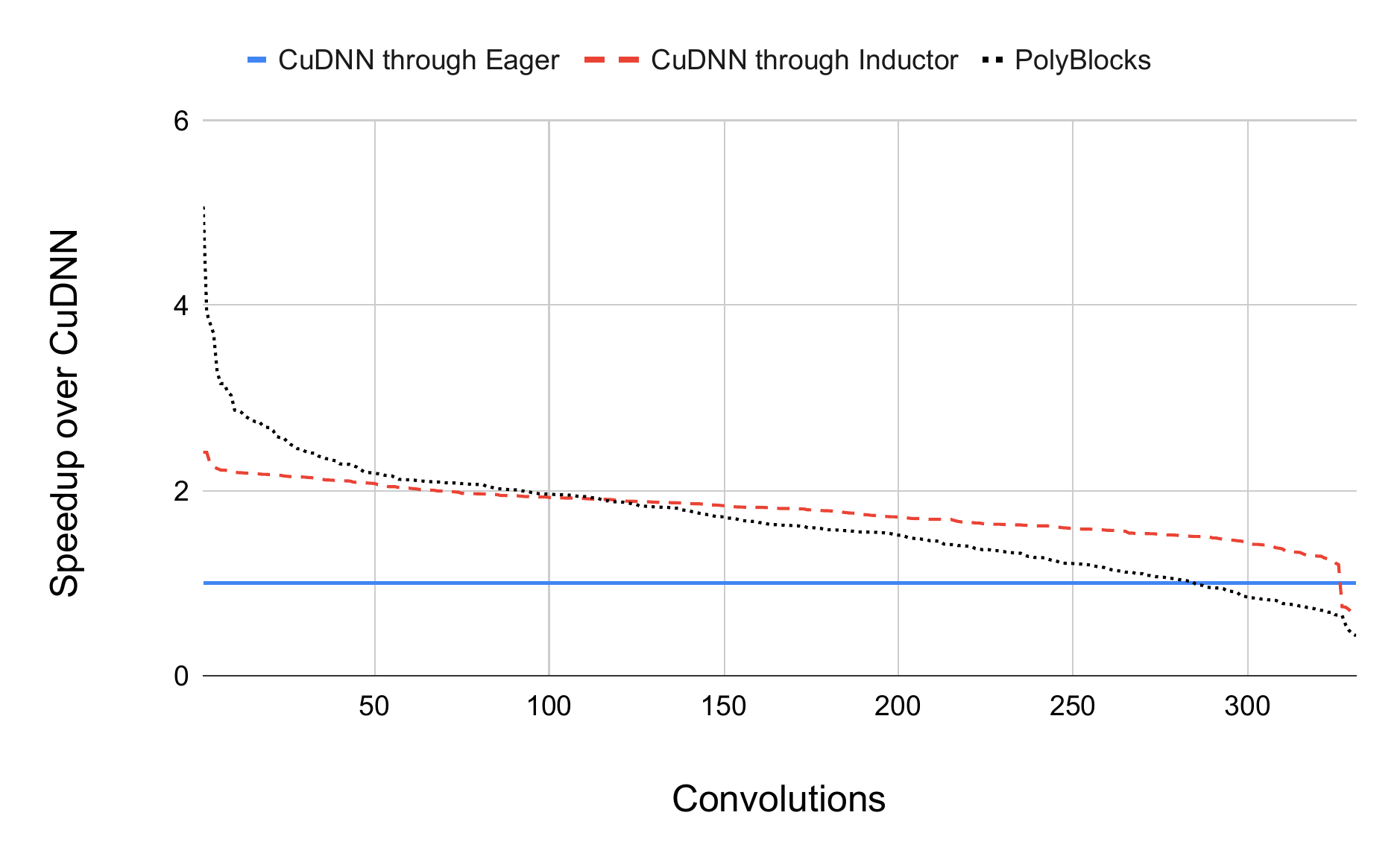} }
	\vskip -6pt
  \caption{Performance of convolution operator code generated fully with
  PolyBlocks on convolutions sourced from various AI models. cuDNN 9.10.2, PyTorch 2.9.0+cu128.
	\label{fig:conv-perf}}
  \vskip -12pt
\end{figure}

Figure~\ref{fig:conv-perf} compares the performance of PolyBlocks with that of
CuDNN on hundreds of convolutions drawn from a large number of vision
models, including those evaluated above.  For convolutions, both torch-eager and
inductor use cuDNN. Inductor uses triton for pointwise kernels fusing any
surrounding ones, while eager uses cuDNN.  The results show that the performance
obtained by code generation through the composition of transformations in
PolyBlocks is competitive with that of CuDNN in a large number of cases. For
several cases (nearly 50), it beats the library by more than 2x.

\begin{figure}[htbp]
	\includegraphics[width=0.60\linewidth]{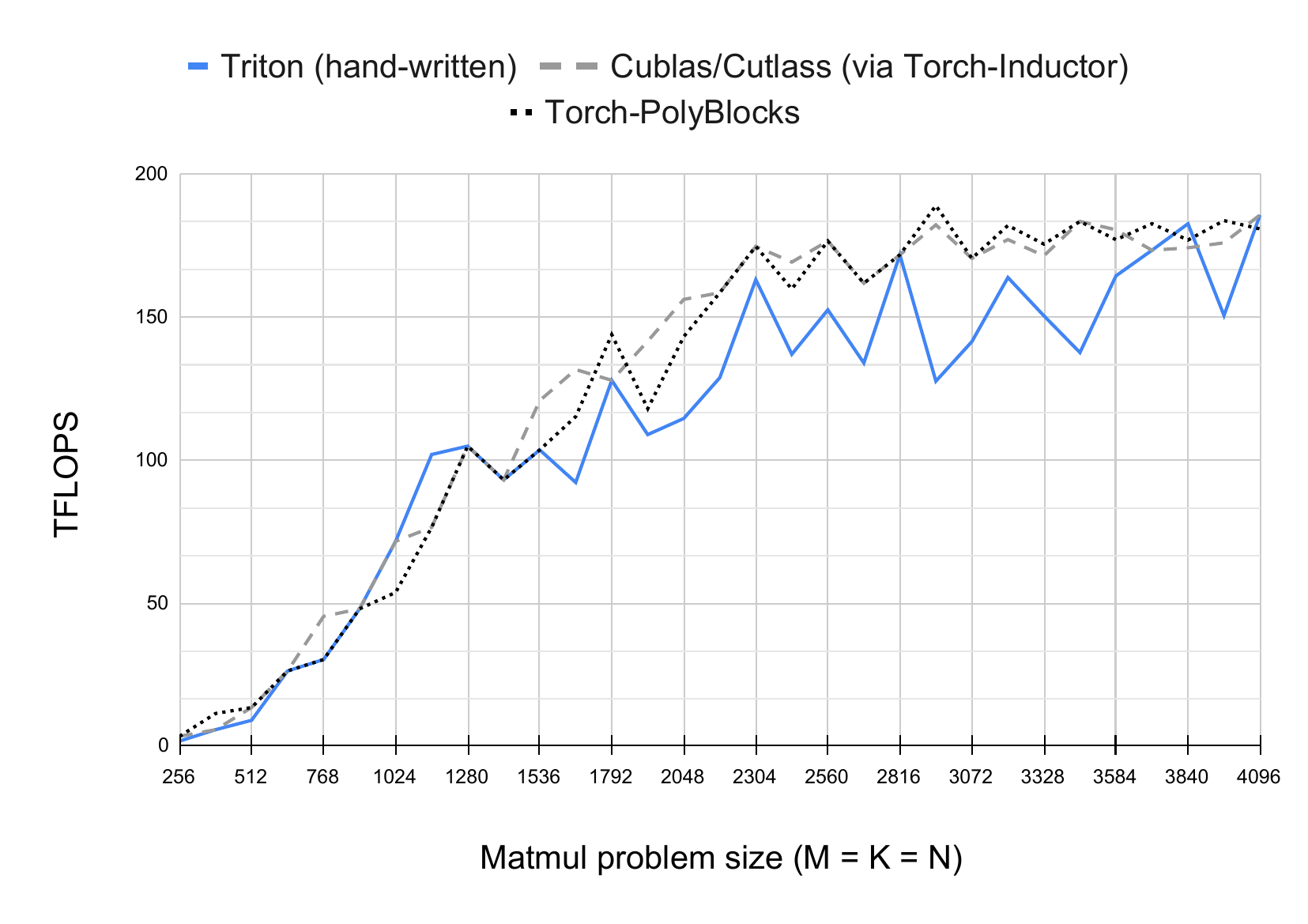}
	\vskip -12pt
	\caption{Performance comparison: CuBLAS, Triton, and PyTorch-PolyBlocks on a single
	matmul on NVIDIA A100. PyTorch 2.7.1. OpenAI Triton 3.3.1. CuBLAS 12.6.4.1.\label{fig:matmul-perf}}
  \vskip -10pt
\end{figure}

Similarly, we compare performance on a single matmul operator for various sizes
in Figure~\ref{fig:matmul-perf}. Here, we compare with Triton and CuBLAS,
representative of state-of-the-art mid- and low-level approaches known to reach
close to machine peak performance. We obtain performance very close to CuBLAS'
on average.

\begin{table}[htbp]
\caption{Attention layer performance for configurations drawn from various
models. RR-fusion is reduce-reduce fusion.  Torch 2.9.0, CUDA 12.8. NVIDIA
A100. Execution times are in milliseconds. \label{tab:attention-perf}}
\vskip -5pt
{\footnotesize
  \begin{tabularx}{\linewidth}{p{0.34\linewidth}p{0.05\linewidth}p{0.07\linewidth}p{0.06\linewidth}p{0.08\linewidth}p{0.06\linewidth}p{0.07\linewidth}p{0.07\linewidth}}
    \toprule
		\multirow{2}{0.75\linewidth}{\textrm{Problem size (batch,
		num-heads, sequence length, dimensionality of head)}}
    & \multicolumn{3}{c}{PolyBlocks} & Inductor &
    \multicolumn{3}{>{\centering\arraybackslash}p{0.27\linewidth}}{PolyBlocks speedup over Inductor} \\
    \cmidrule(lr){2-4} \cmidrule(lr){6-8}  & \textrm{No fusion} &
    \textrm{Only rr-fusion} & \textrm{rr+wmma fusion} & & \textrm{No fusion} &
    \textrm{Only rr-fusion} & \textrm{rr+wmma fusion} \\
\midrule
B=1, H=16, N=512, D=64 & 0.0671 & 0.0516 & 0.0253 & 0.0174 & 0.26 & 0.34 & 0.69 \\
B=1, H=16, N=2048, D=64 & 1.7821 & 0.4770 & 0.2002 & 0.1266 & 0.07 & 0.27 & 0.63 \\
B=1, H=16, N=8192, D=64 & 487.12 & 6.2722 & 2.4359 & 1.5591 & 0.00 & 0.25 &
0.64 \\
B=1, H=16, N=512, D=128 & 0.0854 & 0.1124 & 0.0417 & 0.0449 & 0.53 & 0.40 & 1.08 \\
B=1, H=16, N=2048, D=128 & 1.8915 & 1.0164 & 0.3024 & 0.2296 & 0.12 & 0.23 & 0.76 \\
B=1, H=16, N=8192, D=128 & 487.82 & 13.016 & 3.6674 & 2.7657 & 0.01 & 0.21 &
0.75 \\
DPT-large B=1, H=16, N=577, D=64 & 0.2059 & 0.1013 & 0.0446 & 0.0236 & 0.11 &
0.23
& 0.53 \\
GPT2 B=1, H=12, N=10, D=64 & 0.0168 & 0.0078 & 0.0078 & 0.0082 & 0.49 & 1.06 & 1.06
\\
YOLOs B=1, H=6, N=297, D=64 & 0.0329 & 0.0248 & 0.0185 & 0.0149 & 0.45 & 0.60 &
0.81 \\
RoBERTa B=1, H=16, N=14, D=64 & 0.0151 & 0.0081 & 0.0081 & 0.0083 & 0.55 & 1.03 &
1.03 \\
DistilBERT B=1, H=12, N=13, D=64 & 0.0180 & 0.0081 & 0.0081 & 0.0083 & 0.46 & 1.02
& 1.02 \\
QueryWellFormed B=1, H=80, N=14, D=64 & 0.0157 & 0.0084 & 0.0084 & 0.0085 & 0.54 &
1.01 & 1.01 \\
table\-transf. B=1, H=8, N=500, D=32 & 0.0424 & 0.0402 & 0.0165 & 0.0155 &
0.37 & 0.38 & 0.84 \\
table\_transf. B=1, H=8, N=15, D=32 & 0.0144 & 0.0073 & 0.0073 & 0.0085 &
0.45 & 0.89 & 0.89 \\
owl-vit B=1, H=16, N=577, D=64 & 0.2059 & 0.1013 & 0.0446 & 0.0236 & 0.11 &
0.23 & 0.53 \\
owl-vit B=1, H=16, N=16, D=64 & 0.0142 & 0.0077 & 0.0074 & 0.0083 & 0.58 &
1.07 & 1.13 \\
xlm-roberta B=1, H=12, N=15, D=64 & 0.0154 & 0.0081 & 0.0081 & 0.0082 & 0.54 &
1.02 & 1.02 \\
gte-attn B=1, H=48, N=10, D=64 & 0.0173 & 0.0081 & 0.0081 & 0.0084 & 0.49 & 1.04
&
1.04 \\
mpT7b B=1, H=32, N=1024, D=128 & 0.9601 & 0.5247 & 0.1709 & 0.1235 & 0.13 & 0.24 & 0.72 \\
\bottomrule
\end{tabularx}
}
\vskip -10pt
\end{table}

\paragraph{Performance on the attention layer}
Table~\ref{tab:attention-perf} shows the impact of optimizations described
in Section~\ref{sec:fused-attention} on various attention layer configurations
drawn from transformer-based models. Nsight Systems profile confirms that Torch
Inductor uses hand-written flash attention kernels onto which it maps the
attention layer. The results show the automatic optimization cover significant
ground making the generated code's performance competitive with the
state-of-the-art.


\subsection{Impact of key optimizations: ablation study}
\label{sec:ablation-study}

\begin{table}[tbp]
  \footnotesize
\centering \caption{Torch-PolyBlocks performance impact of key optimizations
  isolated. On the A100.} \label{tab:ablation} \vskip -5pt
\begin{tabularx}{\linewidth}{l *{9}{>{\RaggedLeft\arraybackslash}X}}
\toprule
\textbf{Model} & \multicolumn{5}{c}{\bf Execution times (milliseconds)} & \multicolumn{4}{c}{\bf Speedup (with all opts) over} \\
\cmidrule(lr){2-6}
\cmidrule(lr){7-10}
& \textbf{w/o tensor core} &
\textbf{w/o fusion} &
\textbf{w/o rr fusion} &
\textbf{w/o wmma fusion} &
\textbf{All opts} &
\textbf{no-tensor-core} &
\textbf{no-fusion} &
\textbf{no-rr-fusion} &
\textbf{no-wmma-fusion} \\
\midrule
AlexNet                                 & 2.11     & 0.35    & 0.28     & 0.28    & 0.28    & 7.57   & 1.24   & 1.00   & 1.00 \\
ConvNext                                & 12.32    & 2.83    & 0.94     & 0.93    & 0.93    & 13.20  & 3.03   & 1.01   & 1.00 \\
DenseNet                                & 9.14     & 5.31    & 2.26     & 2.27    & 2.23    & 4.11   & 2.39   & 1.02   & 1.02 \\
EfficientNet2                           & 11.82    & 4.87    & 1.42     & 1.42    & 1.40    & 8.43   & 3.47   & 1.01   & 1.01 \\
GoogLeNet                               & 4.19     & 2.02    & 0.57     & 0.57    & 0.55    & 7.65   & 3.68   & 1.03   & 1.04 \\
Inception                               & 9.11     & 3.38    & 1.01     & 1.01    & 0.99    & 9.20   & 3.41   & 1.02   & 1.02 \\
MNasNet                                 & 2.00     & 1.96    & 0.43     & 0.43    & 0.42    & 4.81   & 4.72   & 1.04   & 1.04 \\
MobileNet3                              & 1.65     & 2.10    & 0.46     & 0.46    & 0.45    & 3.67   & 4.68   & 1.03   & 1.03 \\
ShuffleNet                              & 2.34     & 2.14    & 0.54     & 0.55    & 0.53    & 4.44   & 4.06   & 1.02   & 1.03 \\
SqueezeNet                              & 1.40     & 0.49    & 0.18     & 0.18    & 0.18    & 7.74   & 2.73   & 1.01   & 1.01 \\
HRNet                                   & 21.06    & 12.19   & 4.41     & 4.40    & 4.07    & 5.17   & 2.99   & 1.08   & 1.08 \\
ResNet50                                & 7.24     & 2.03    & 0.59     & 0.59    & 0.59    & 12.34  & 3.46   & 1.01   & 1.01 \\
VGG19                                   & 19.63    & 1.36    & 1.00     & 1.00    & 0.98    & 20.12  & 1.40   & 1.02   & 1.02 \\
UNet                                    & 192.8    & 5.12    & 2.15     & 2.15    & 2.10    & 91.83  & 2.44   & 1.02   & 1.02 \\
VIT                                     & 59.96    & 5.24    & 2.38     & 2.21    & 2.19    & 27.33  & 2.39   & 1.09   & 1.01 \\
GPT2                                    & 114.43   & 15.29   & 6.43     & 7.96    & 7.93    & 14.43  & 1.93   & 0.81   & 1.00 \\
YOLOS                                   & 86.03    & 3.99    & 1.36     & 1.32    & 1.22    & 70.77  & 3.28   & 1.12   & 1.08 \\
RoBERTa                                 & 202.9    & 12.61   & 5.89     & 5.88    & 6.41    & 31.66  & 1.97   & 0.92   & 0.92 \\
DistilBERT                              & 23.28    & 2.25    & 0.95     & 1.15    & 1.13    & 20.64  & 1.99   & 0.85   & 1.02 \\
Query wellformedness score              & 46.52    & 4.51    & 1.83     & 1.86    & 1.94    & 23.93  & 2.32   & 0.94   & 0.96 \\
Stable diffusion turbo unet             & 462.0    & 43.72   & 29.01    & 17.14   & 14.6    & 31.74  & 3.00   & 1.99   & 1.18 \\
Stable diffusion XL1 base unet          & 7181     & 720.3   & 1701     & 156.8   & 137.9   & 52.09  & 5.23   & 12.34  & 1.14 \\
Stable diffusion XL1 refiner unet       & 7422     & 1052.6  & 3889     & 176.0   & 155.0   & 47.88  & 6.79   & 25.09  & 1.14 \\
Flux transformer                        & 33438    & 2168.9  & 13510    & 696.7   & 451.3   & 74.09  & 4.81   & 29.93  & 1.54 \\
Table Transformer                       & 66.08    & 5.37    & 2.21     & 2.27    & 2.37    & 27.83  & 2.26   & 0.93   & 0.95 \\
OWL VIT Base Patch                      & 138.76   & 10.35   & 4.88     & 3.92    & 3.16    & 43.93  & 3.28  & 1.55   & 1.24 \\
DPT Large                               & 496.14   & 0.00    & 47.72    & 43.59   & 42.12   & 11.78  & 0.00   & 1.13   & 1.03 \\
XLM Roberta                             & 187.39   & 7.72    & 3.30     & 3.44    & 3.44    & 54.46  & 2.24   & 0.96   & 1.00 \\
GTE                                     & 14.72    & 3.65    & 1.30     & 1.75    & 0.94    & 15.63  & 3.87   & 1.38   & 1.86 \\
MPT-7B                                  & 516.74   & 22.87   & 18.95    & 18.44   & 18.45   & 28.00  & 1.24   & 1.03   & 1.00 \\
\midrule
{\bf Geomean speedup}                   &          &         &          &         &         & 17.36  & 2.87   & 1.42   & 1.07 \\
\bottomrule
\end{tabularx}
\vskip -10pt
\end{table}

We now isolate the individual impact of certain optimizations and related
hardware features. Table~\ref{tab:ablation} data shows how well-optimized code
behaves with and without tensor core capability, cross-operator fusion, and
reduce-reduce fusion. Tensor cores deliver a geomean speedup of 17x. A100 has
a peak 312 mixed precision TFLOPS using tensor cores versus 19.5~TFLOPS
using fp32. Also, fp16 data types with fusion lead to 2x as
fast movement of tensors to and from global memory. Higher than 16-20x
speedups in some cases can be considered anomalies due to CUDA core-targeted
code not being optimized as well by the remaining passes of our pipeline.
Separately, cross-operator fusion delivers a 2.87x improvement for otherwise fully
optimized code, showing the importance of reducing global memory roundtrips and
hiding inherently memory-bandwidth bound operations by exploiting on-chip shared
memory or register reuse. The table also shows the impact reduce-reduce and wmma
fusion have for workloads with the attention layer, isolating the impact of
Section~\ref{sec:fused-attention} on full workloads.

\section{Related Work}
\label{sec:related-work}

The work on BLIS~\cite{vanzee2015toms,low2016toms} showed how peak hardware
performance could be achieved for matrix-matrix multiplication on CPUs in a more
modular and reusable way, with cost models and additional tooling. Early
evidence of achieving close to peak performance on par with the best
hand-written libraries completely using MLIR code generation for CPUs and GPUs,
albeit in a preliminary setting, was soon demonstrated~\cite{bondhugula20arxiv,
katel22cc}.

There have been a large number of deep learning compiler efforts in the past
decade. Many of these were comprehensively evaluated by Ansel et
al.~\cite{pytorch24asplos}.
Torch Inductor was shown to be vastly better both in compilation coverage and in
performance on GPUs and CPUs~\cite{pytorch24asplos}. Prior work compared
there included nvFuser~\cite{sarofeen2022nvfuser},
NNC~\cite{zolotukhin2021nnc}, XLA~\cite{pytorchxla2023},
ONNXRT~\cite{onnxruntime2021}, TVM~\cite{chen18osdi}, and
Hidet~\cite{ding23asplos}. A good part of our evaluation was thus focused on
using Inductor as a reference for the state-of-the-art for PyTorch.

\textbf{High-level AI compilers:} XLA~\cite{xla} and Torch
Inductor~\cite{pytorch24asplos} are the prominent production-strength
high-level compilers available as part of the respective official repositories.
While XLA serves JAX and TensorFlow, Inductor is for PyTorch and potentially
other frameworks that can be rewritten to torch graphs.  They provide high
compilation coverage and the desired level of automation.  Models that work
with eager execution typically work out of the box and unmodified with JIT
compilation turned on.  Both XLA and Inductor differ vastly from PolyBlocks in the
intermediate representation used to perform transformations. While Inductor
uses its own IR abstractions (torch fx IR with aten operations) before relying
on Triton for GPU compilation, XLA uses the HLO~\cite{xla-hlo} representation
for its high-level transformations before using Triton and LLVM emitters for
CPUs/GPUs or LLO for TPUs~\cite{xla-tpu}. Both of these compilers heavily rely
on vendor libraries or other open-source kernels for compute-intensive
operators on GPUs: CuDNN for convolutions, CuBLAS for matmuls, and various
flash attention kernels for the attention layer. In contrast, PolyBlocks is fully
code generating and the benefits of this approach were discussed in more
detail in Section~\ref{sec:design-choices}. While the work on
PyTorch-2~\cite{pytorch24asplos} was a milestone result in compilation and
auto-parallelization, the resulting infrastructure is not easily reusable for
other programming frameworks and hardware due to its choice of multiple IRs and
reliance on libraries to perform some of the heavy lifting.  The optimizations
we described in Sections \ref{sec:fusion}, \ref{sec:conv-mapping}, and
\ref{sec:fused-attention} are not only not performed in Inductor and XLA, but
are difficult to achieve due to the nature and level of IR abstractions available.
While our fusion approach is similar in spirit to that of
Halide~\cite{halide,mullapudi2016siggraph} in allowing redundant computation
and the choice of depth to compute at, it differs in its choice of intermediate
representation, the machinery used to automatically compute slices, cost
models, and applicability to n-dimensional affine nests (perfect or imperfect) as opposed
to only image processing computations targeted by Halide.


\textbf{Mid-level compilers:} In recent years, mid-level programming frameworks such
as Triton~\cite{triton19tillet,triton-web} have become popular and inspired other
similar frameworks~\cite{pallas, helion, spector2024arxiv}. Such
frameworks make programmers use tile-level abstractions. Since programmers
still perform the tiling and fusion themselves,  considerable effort is needed
to express medium to large graphs of computations. On the other hand,
high-level frameworks like Inductor, XLA, or PolyBlocks would perform tiling and
fusion transparently and automatically.  The optimizations we described in
Sections \ref{sec:fusion}, \ref{sec:conv-mapping}, \ref{sec:fused-attention}, as
well as quantization (Appendix~\ref{sec:quantization}) are not performed in
Triton, and are expected to be performed by the programmer with considerable
effort.  Mid-level frameworks are thus more suitable as kernel-writing
languages to substitute out compute-intensive parts of a graph that were
otherwise a bottleneck with high-level approaches. We expect these approaches
to co-exist with both lower- and high-level approaches. While
FlexAttention~\cite{dong24arxiv} is able to generate various configurations of
attention, it is restricted to that class of computations. Our approach is more
general, but does not yet support the entire cross product of attention
configurations with a few aspects left for future implementation in the relevant
passes that perform the reduce-reduce fusion described in
Section~\ref{sec:fused-attention}.

IREE~\cite{iree2019} is an MLIR-based deep learning compiler effort with similar
goals on certain aspects, such as code generation, as PolyBlocks. However, it
differs significantly from PolyBlocks in its choice of dialects and types to
perform mid-level optimizations. Based on the evaluation provided in
Appendix~\ref{sec:iree-perf}, it does not provide the level of transformation
capability or model coverage as Inductor, XLA, or PolyBlocks for GPUs, with certain
basic optimization techniques missing.

{\bf Polyhedral works:} Several works in the polyhedral compilation literature
developed frameworks to perform optimizations for parallelism and locality in
both general-purpose and domain-specific settings~\cite{bondhugula08pldi,
ppcg, mullapudi2015asplos, grosser15toplas, vasilache18tc, elango2018mapl,
zinenko18cc, bhaskaracharya20arxiv, zhao24tocs}. These works use
the mathematical notion of domains and schedules, as integer sets and affine
functions (or relations) respectively, to represent, compose, and apply
transformations. IR materializes only after the polyhedral AST generation step.
The approach of PolyBlocks is drastically different from being built like a
standard compiler pass pipeline where optimizations are performed using both
affine analysis and SSA in smaller steps.  This allows for more modular and
scalable compiler development and testing, better interaction with SSA concepts,
and significantly faster compile times. Separately, the fusion approach we have
used is fundamentally different from polyhedral loop fusion approaches, which
are not slicing based.  To our knowledge, the work of Zhao et
al.~\cite{zhao24tocs} is among the most sophisticated works on tiling and fusion
using the polyhedral approach. This approach improves upon or subsumes various
prior approaches by combining tiling and fusion with backward slicing. However,
the approach uses full-fledged integer set computations and is expensive in
comparison to ours. It was also evaluated on smaller benchmarks compared to the
ones we used here (cf.  Section~\ref{sec:ablation-study}). Scalability would
thus be a challenge to inputs with hundreds to thousands of nests. In contrast,
our approach relies on both SSA and MLIR's dependence constraints in its flat
value constraints form to construct slices easily for the domain of affine nests
we encountered.

\section{Conclusions}
\label{sec:conclusions}
We presented the design and implementation of PolyBlocks, a modular and reusable
MLIR-based compiler infrastructure for AI programming frameworks and accelerator
chips.  PolyBlocks uses sequences of pass pipelines that perform transformations on
loop nests, static single assignment (SSA), and multidimensional memory accesses
primarily using lightweight affine analysis.  Simple transformations are
composed together in specialized ways to realize advanced transformations
by the use of analytical cost models and heuristics. The developed passes
perform transformations related to multi-level tiling, fusion, on-chip
scratchpad usage, mapping matmuls and convolutions to matrix units, fusing
attention layer computation, and several other well-known transformations for
parallelism and locality specialized for dense tensor computations.
Experimental results from evaluating PolyBlocks-based compiler backends for PyTorch
and JAX for NVIDIA GPUs show that it is able to match or outperform Torch
Inductor and XLA by significant margins in several cases, despite the latter
relying on a combination of libraries and code generation. For individual
compute-intensive operators, PolyBlocks-based compiler code is competitive with
vendor-tuned libraries or kernels written in mid-level frameworks such as
Triton. The design and architecture of PolyBlocks provide a way to quickly build
new compilers for computations expressed in frameworks like PyTorch and JAX
targeting new chips with specialized features.

\begin{acks}
We would like to acknowledge community contributions to various open-source
projects that PolyBlocks uses, including MLIR, torch-mlir, and LLVM. We would
also like to acknowledge the contributions of Prathamesh Tagore and Vinayaka
Bandishti at the time they were employed at Polymage Labs.

\end{acks}

\bibliographystyle{ACM-Reference-Format}
\bibliography{references}

\newpage
\appendix

\section{Supplemental Material for ``PolyBlocks: A Compiler Infrastructure for AI
Chips and Programming Frameworks''}

\subsection{Quantization} \label{sec:quantization}

PolyBlocks supports PyTorch models quantized using PyTorch~2 export
quantization~\cite{pt2eq} and TensorFlow models quantized using
TensorFlow-2.x-Quantization-Toolkit~\cite{tf-quant}. Optimizations for uniform
symmetric int8-quantized models are currently supported. IR extracted from
models quantized using both workflows contains quantize and dequantize stub
operations without enclosing any compute operations. At the {\it mhlo} dialect
level, in the S1 stage of the pipeline~\ref{fig:polyblocks-components},
transformation passes reposition compute-intensive operations such as matmuls
and convolutions between the quantize and dequantize operations, enabling these
operations to be executed in {\it int8} format. All optimization that works on
non-quantized models, like fusion, vectorization, and matrix-unit mapping, works
out of the box for quantized models. An example showing the fusion of nearly
15-20 operators into the output of a quantized tensor-core mapped convolution is
shown in Figure~\ref{fig:fused-quant-nest}. This example shows tiling, fusion,
mapping to tensor cores, and vectorization of on-chip to/from global memory data
movement code.

\begin{figure}[htbp]
\includegraphics[width=\linewidth]{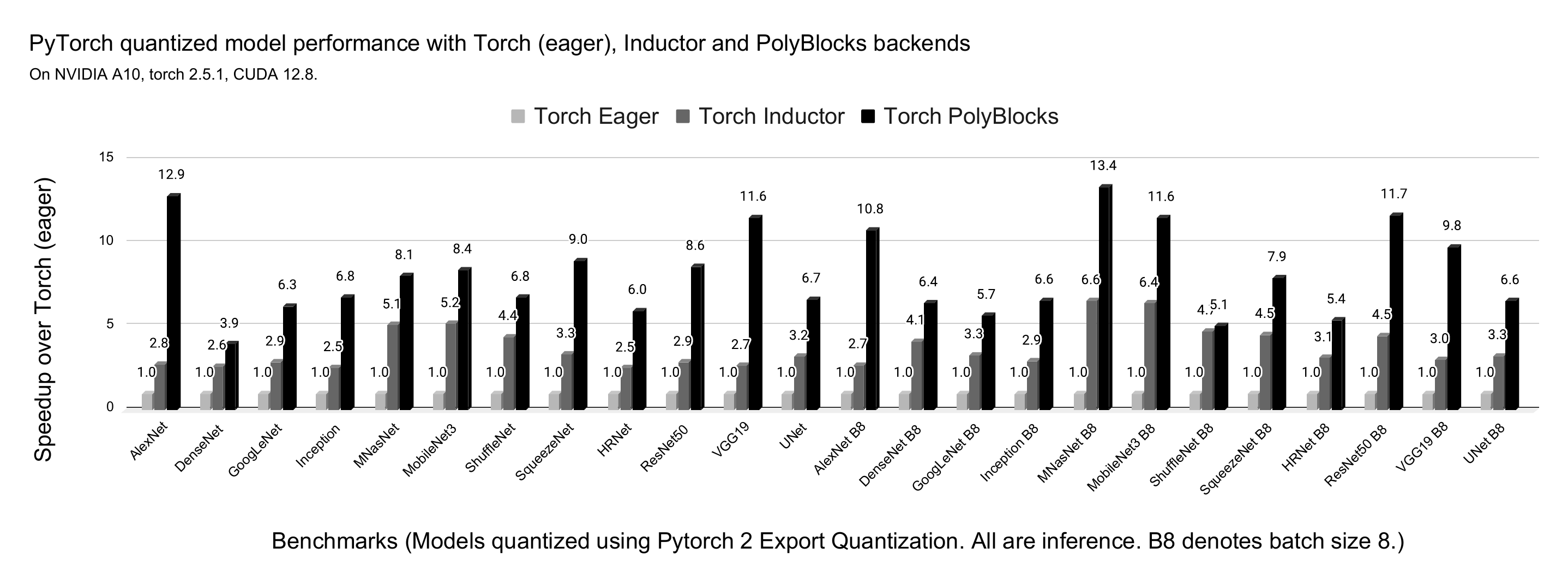}
  \caption{
  PyTorch i8 quantized performance with various backends: eager, Inductor, TensorRT, and PolyBlocks. NVIDIA A10.
  Torch 2.5.1, torch-tensorrt-2.5.0, TensorRT 10.3.0.
\label{fig:quantized-perf-a10}}
\end{figure}

\begin{figure}[htbp]
\includegraphics[width=\linewidth]{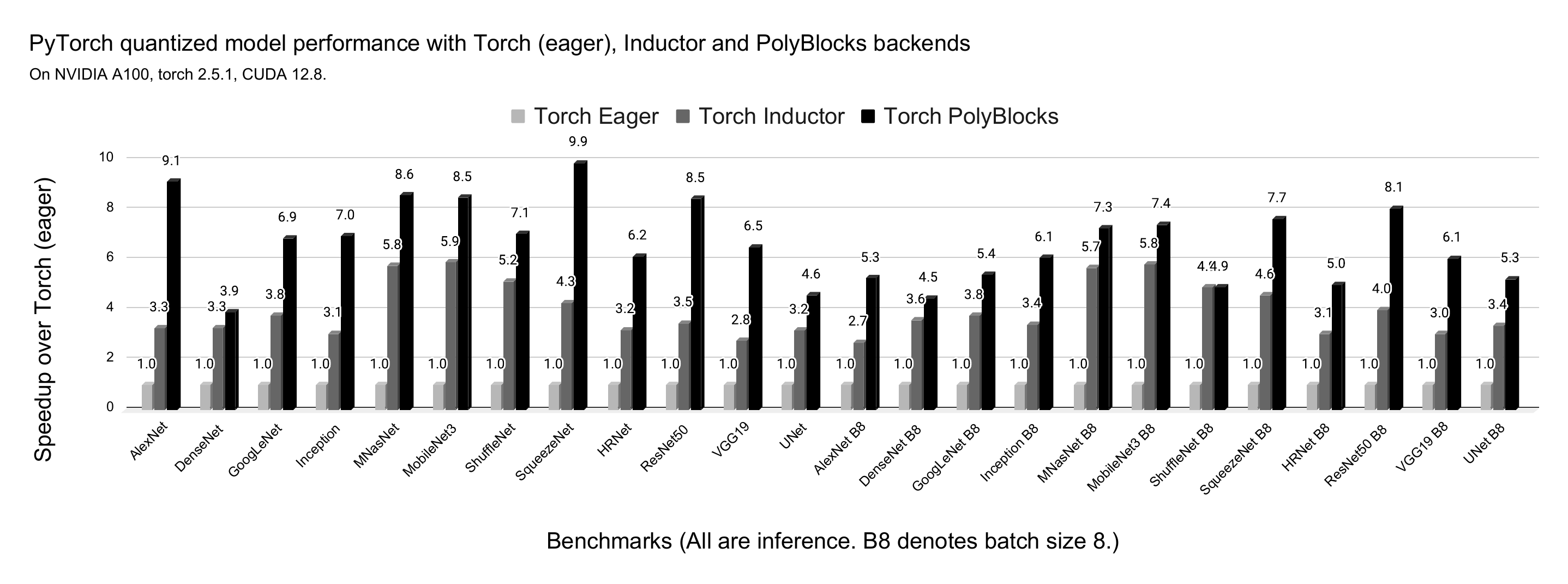}
  \caption{
  PyTorch i8 quantized performance with various backends: eager, Inductor, TensorRT, and PolyBlocks. NVIDIA A100.
  Torch 2.5.1, torch-tensorrt-2.5.0, TensorRT 10.3.0.
\label{fig:quantized-perf-a100}}
\end{figure}

\begin{figure}[htbp]
\includegraphics[width=\linewidth]{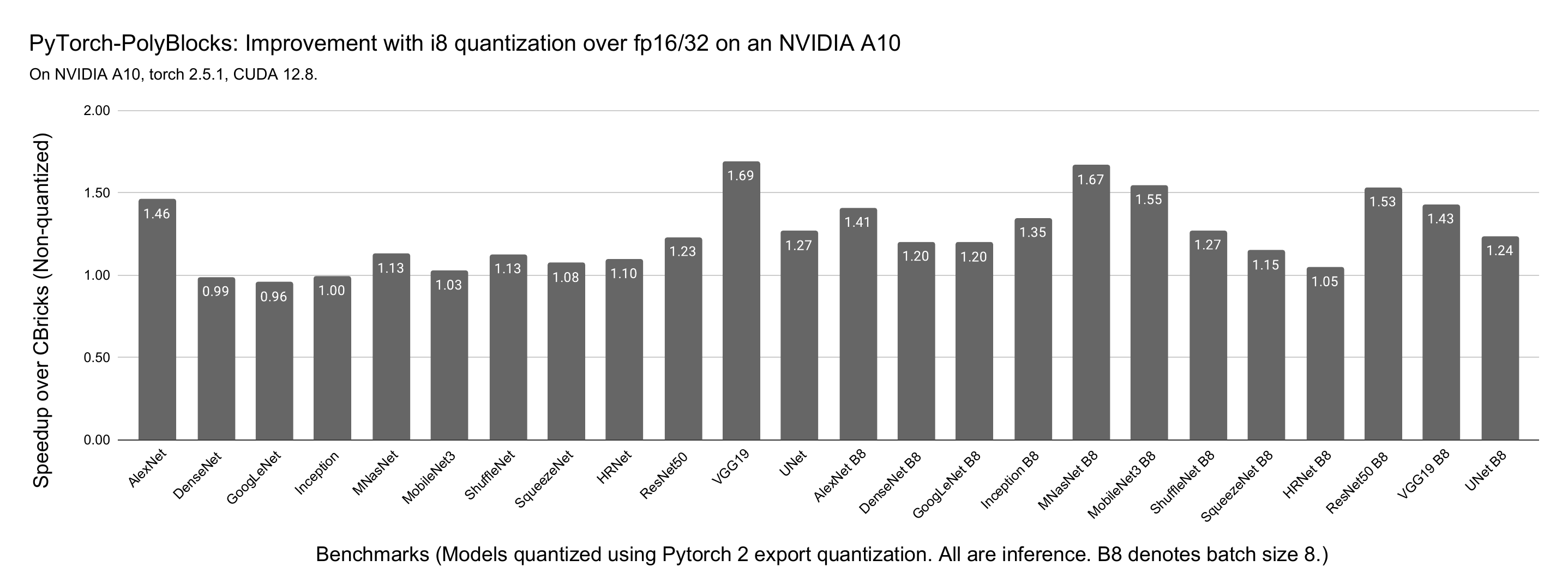}
  \caption{
  PyTorch-PolyBlocks i8 quantized performance vs fp16/fp32: batch size 8 on the NVIDIA A10.
\label{fig:quantized-perf-i8-over-fp16}}
\end{figure}

Figures~\ref{fig:quantized-perf-a10} and \ref{fig:quantized-perf-a100} compare
the performance of PolyBlocks with quantization to 8-bit integers. Torch-eager and
Inductor have limited support for optimization in the presence of quantization.
We thus observe dramatic speedups when using PolyBlocks. Furthermore,
Figure~\ref{fig:quantized-perf-i8-over-fp16} demonstrates the benefit with int8
quantization over fp16/fp32, putting the performance with quantization in
perspective with other reported performances in Section~\ref{sec:evaluation}.

\begin{figure}[htbp]
\begin{lstlisting}[language=MLIR,basicstyle=\fourpointsize\ttfamily]
        // Tensor core mapped compute on a tile.
        ...
        %520 = gpu.subgroup_mma_compute %508, %519, %arg166 {b_transpose} :
           !gpu.mma_matrix<16x16xi8, "AOp">, !gpu.mma_matrix<16x16xi8, "BOp"> ->
           !gpu.mma_matrix<16x16xi32, "COp">
        %521 = gpu.subgroup_mma_compute %511, %519, %arg167 {b_transpose} :
           !gpu.mma_matrix<16x16xi8, "AOp">, !gpu.mma_matrix<16x16xi8, "BOp"> ->
           !gpu.mma_matrix<16x16xi32, "COp">
        affine.yield %510, %512, %514, %515, %517, %518, %520, %521 :
           !gpu.mma_matrix<16x16xi32, "COp">, !gpu.mma_matrix<16x16xi32, "COp">,
           !gpu.mma_matrix<16x16xi32, "COp">, !gpu.mma_matrix<16x16xi32, "COp">,
           !gpu.mma_matrix<16x16xi32, "COp">, !gpu.mma_matrix<16x16xi32, "COp">,
           !gpu.mma_matrix<16x16xi32, "COp">, !gpu.mma_matrix<16x16xi32, "COp">
      }
      gpu.barrier
      gpu.subgroup_mma_store_matrix %506#0, %view_574[%494, %c0] {leadDimension = 68 : index} : !gpu.mma_matrix<16x16xi32, "COp">, memref<128x68xi32, 3>
      gpu.subgroup_mma_store_matrix %506#1, %view_574[%496, %c0] {leadDimension = 68 : index} : !gpu.mma_matrix<16x16xi32, "COp">, memref<128x68xi32, 3>
      gpu.subgroup_mma_store_matrix %506#2, %view_574[%494, %c16] {leadDimension = 68 : index} : !gpu.mma_matrix<16x16xi32, "COp">, memref<128x68xi32, 3>
      gpu.subgroup_mma_store_matrix %506#3, %view_574[%496, %c16] {leadDimension = 68 : index} : !gpu.mma_matrix<16x16xi32, "COp">, memref<128x68xi32, 3>
      gpu.subgroup_mma_store_matrix %506#4, %view_574[%494, %c32] {leadDimension = 68 : index} : !gpu.mma_matrix<16x16xi32, "COp">, memref<128x68xi32, 3>
      gpu.subgroup_mma_store_matrix %506#5, %view_574[%496, %c32] {leadDimension = 68 : index} : !gpu.mma_matrix<16x16xi32, "COp">, memref<128x68xi32, 3>
      gpu.subgroup_mma_store_matrix %506#6, %view_574[%494, %c48] {leadDimension = 68 : index} : !gpu.mma_matrix<16x16xi32, "COp">, memref<128x68xi32, 3>
      gpu.subgroup_mma_store_matrix %506#7, %view_574[%496, %c48] {leadDimension = 68 : index} : !gpu.mma_matrix<16x16xi32, "COp">, memref<128x68xi32, 3>
    }
    affine.parallel (%arg158, %arg159) = (0, 0) to (128, 16) {
      %494 = affine.load %493[%arg158, %arg159] : memref<128x17xvector<4xi32>, 3>
      %495 = arith.sitofp %494 : vector<4xi32> to vector<4xf32>
      %496 = affine.load %10[%arg159] : memref<16xvector<4xf32>>
      %497 = arith.mulf %495, %496 : vector<4xf32>
      %498 = affine.load %11[%arg159] : memref<16xvector<4xf32>>
      %499 = arith.addf %497, %498 : vector<4xf32>
      %500 = arith.divf %499, %cst_35 : vector<4xf32>
      %501 = math.roundeven %500 : vector<4xf32>
      %502 = arith.maximumf %501, %cst_6 : vector<4xf32>
      %503 = arith.minimumf %502, %cst_5 : vector<4xf32>
      %504 = arith.fptosi %503 : vector<4xf32> to vector<4xi8>
      %505 = arith.sitofp %504 : vector<4xi8> to vector<4xf32>
      %506 = affine.load %12[%arg159] : memref<16xvector<4xf32>>
      %507 = arith.truncf %cst_61 : f64 to f32
      %508 = vector.splat %507 : vector<4xf32>
      %509 = arith.addf %506, %508 : vector<4xf32>
      %510 = vector.extract %509[0] : vector<4xf32>
      %511 = math.sqrt %510 : f32
      %512 = vector.insert %511, %509 [0] : f32 into vector<4xf32>
      %513 = vector.extract %509[1] : vector<4xf32>
      %514 = math.sqrt %513 : f32
      %515 = vector.insert %514, %512 [1] : f32 into vector<4xf32>
      %516 = vector.extract %509[2] : vector<4xf32>
      %517 = math.sqrt %516 : f32 loc(#loc526
      %518 = vector.insert %517, %515 [2] : f32 into vector<4xf32>
      %519 = vector.extract %509[3] : vector<4xf32>
      %520 = math.sqrt %519 : f32
      %521 = vector.insert %520, %518 [3] : f32 into vector<4xf32>
      %522 = arith.mulf %505, %cst_35 : vector<4xf32>
      %523 = arith.divf %cst_2, %521 : vector<4xf32>
      %524 = affine.load %13[%arg159] : memref<16xvector<4xf32>>
      %525 = arith.subf %522, %524 : vector<4xf32>
      %526 = arith.mulf %525, %523 : vector<4xf32>
      %527 = affine.load %14[%arg159] : memref<16xvector<4xf32>>
      %528 = arith.mulf %526, %527 : vector<4xf32>
      %529 = affine.load %15[%arg159] : memref<16xvector<4xf32>>
      %530 = arith.addf %528, %529 : vector<4xf32>
      %531 = arith.cmpf ugt, %530, %cst_0 : vector<4xf32>
      %532 = arith.select %531, %530, %cst_0 : vector<4xi1>, vector<4xf32>
      %533 = arith.divf %532, %cst_34 : vector<4xf32>
      %534 = math.roundeven %533 : vector<4xf32>
      %535 = arith.maximumf %534, %cst_6 : vector<4xf32>
      %536 = arith.minimumf %535, %cst_5 : vector<4xf32>
      %537 = arith.fptosi %536 : vector<4xf32> to vector<4xi8>
      affine.store %537, %17[%arg157 floordiv 800, ((%arg158 + %arg157 * 128) mod 102400) floordiv 320,
                        (%arg158 + %arg157 * 128) mod 320, %arg159] : memref<8x320x320x16xvector<4xi8>>
  }
}
\end{lstlisting}
\caption{IR after fusion of de-quantization and quantization at the output of a
convolution operator.\label{fig:fused-quant-nest}}
\end{figure}

\subsection{Overlapping compute with data copy}
\label{sec:compute-copy-overlap}
With the increasing disparity between memory bandwidth and compute throughput,
it is critical to overlap data movement with compute explicitly for many
accelerator chips. Our pass pipeline realizes this through a special pass.  The
pass is powerful enough to overlap data movement at multiple levels of the
memory hierarchy with compute, as is common in modern GPUs. Furthermore, the
pass can overlap multiple stages of data movement with compute. The pass
leverages an affine-loop-shifting transformation to realize the compute-copy
overlap by shifting the data-copy ahead of the compute by the number of stages
determined by a heuristic-based analytical cost model. The cost model considers
various aspects like tile sizes and the amount of available on-chip memory to
determine the number of copy stages to overlap.

After the shifting transformation, the pass maps the data-movement loops to
NVIDIA async-copy operations~\cite{nvidia-ptx-async-copy-operations}. As these
are asynchronous copy operations (executing thread does not wait for the copy to
complete), the pass then inserts necessary synchronization operations to ensure
correctness and completes the transformation. Enabling this transformation plays
a key role in achieving close to peak performance for unit workloads like
matmul, convolutions, and fused-attention.

\subsection{Comparison with IREE}
\label{sec:iree-perf}
We also compare with another MLIR-based code generator, although it has limited
coverage and optimization capability. IREE performance in
Table~\ref{tab:iree-perf} shows that its optimizer is missing significant
transformations including mapping to tensor cores for non-trivial models. IREE's
frontend integration also failed to lower six of the nine models we were able to
try.

\subsubsection{IREE setup}
Details of the IREE version and setup used for benchmarking are provided below.
\texttt{iree-base-compiler, iree-base-runtime, iree-turbine version
3.8.0} through the official PIP package.

\begin{table}[htbp]
%
%
\centering
\captionof{table}{Comparison with IREE. IREE version 3.8.0 (PIP). All execution
times are
in milliseconds.}
\label{tab:iree-perf}
\begin{tabularx}{1\linewidth}{l c c c}
	\toprule
  Workload	& \multicolumn{2}{c}{IREE (milliseconds)} &	PolyBlocks (milliseconds)
  \\
	\cmidrule(lr){2-3}
  & fp32 with AMP  & fp16 & mixed fp16/fp32 \\
	\midrule
	Alexnet batch 8 &	17.92 &	15.21 &	0.36 \\
	VGG19 batch 8 &	225.59 & 189.65	&	2.81 \\
  Unet batch 8 &	1353.88 &	1303.28 &	13.2 \\
	DenseNet batch 8 &	Crash & Crash & 3.49 \\
	Inception batch 8 &	Crash & Crash & 1.85 \\
	Resnet batch 8 &	Crash & Crash & 1.74 \\
	Vit batch 8 & Fails to lower & Fails to lower & 4.51 \\
	HRNet batch 8	 & Fails to lower & Fails to lower & 6.37 \\
	YOLOs	& Fails to lower & Fails to lower & 1.21 \\
	\bottomrule
  Geomean speedup with PolyBlocks & & & 65.5x \\
	\bottomrule
\end{tabularx}
\end{table}

\texttt{IREE compiler flags:}
\begin{verbatim}
  --iree-codegen-llvmgpu-use-igemm
  --iree-codegen-llvmgpu-use-mma-sync
  --iree-codegen-llvmgpu-igemm-pad-convolution
  --iree-codegen-llvmgpu-use-tile-and-fuse-matmul
  --iree-codegen-llvmgpu-use-tile-and-fuse-convolution
  --iree-cuda-target-features=+ptx87 --iree-opt-level=O3
\end{verbatim}

\subsection{PolyBlock kinds and attributes}
\label{sec:polyblocks-kinds}
We classify affine nests into four kinds: (1) pointwise, (2) broadcast, (3)
matmul (which includes convolutions as well), and (4) miscellaneous affine
(`misc'). These polyblocks kinds are marked as attributes on the root affine nest
and then used in some of the passes as hints in guiding to a limited extent some
of the optimization choices. PolyBlock kinds are determined primarily from the
nature of affine access patterns in relation to the surrounding loop induction
variables.

The absence of these hints is guaranteed not to affect correctness, but only
optimizations and performance. As an example, we prevent certain kinds of
fusion from happening into matmul nests that may prevent mapping to matrix units
when available. Strictly speaking, such analysis can be performed on demand with
some additional overhead and code burden. So, the presence of PolyBlocks types can
be thought of as for convenience and to reduce additional detection overhead or
burden.

We also make use of external attributes to convey optimization information and
hints over short distances in the pass pipeline, where distance indicates the
number of passes, and avoid recomputation of certain properties. The target
information is supplied through module attribute information from the driver
(C4 in Section~\ref{sec:polyblocks-stages}).  The root loops of the nests are
marked with the target early in stage S3.

\subsection{Compilation times and model coverage}
\label{sec:compilation-time-coverage}

Some of the largest models evaluated in Section~\ref{sec:evaluation} involve a
few thousand torch aten operators and up to tens of thousands of affine nests in
the unoptimized form at the entry point of S3
(Figure~\ref{fig:polyblocks-components}).  Compilation with PolyBlocks is completed in
tens of seconds to two minutes, with a few models compiling in a few seconds.
For reference, the smallest models such as Alexnet take four seconds to compile
through all five stages of PolyBlocks, with 1.7 seconds in the key optimization
stage S3. The largest models, such as those of the order of Stable Diffusion XL1
refiner Unet take about 10 minutes to compile, with about 7 minutes spent in
stage S3. Most medium-sized models take a few tens of seconds. We thus consider
the infrastructure quite scalable from the compilation time standpoint, although
there are opportunities to optimize compile times.

\begin{figure}[htbp]
\includegraphics[width=\linewidth]{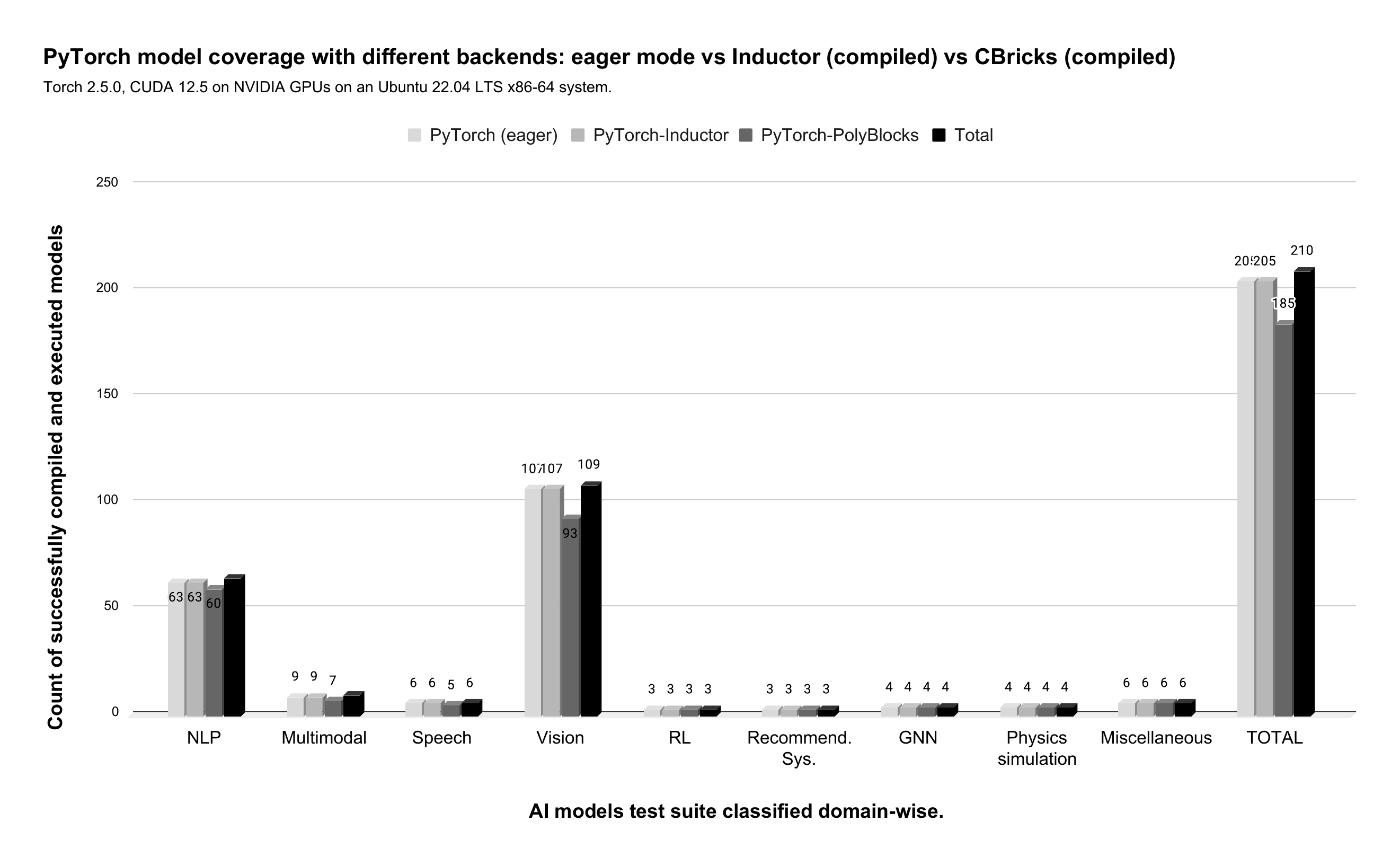}
  \caption{
PyTorch model coverage with PolyBlocks.
\label{fig:pytorch-model-coverage}}
\end{figure}

Figure~\ref{fig:pytorch-model-coverage} shows that PolyBlocks is able to
successfully compile and execute 90\% of the 205 models that were used in the
evaluation in PyTorch-2 work~\cite{pytorch24asplos}, which both torch-eager
and Inductor successfully execute. All 10\% that fail are due to
operator lowering issues in torch-mlir~\cite{torch-mlir} as opposed to core
compile engine issues.

\end{document}